\documentclass[10pt,journal,cspaper,compsoc]{IEEEtran}
\usepackage{verbatim}
\usepackage{graphicx}
\usepackage{subfigure}
\usepackage{comment}
\usepackage{array}
\usepackage{mathtools}
\usepackage{amsmath}
\usepackage{amssymb}
\usepackage{color}
\usepackage{times}
\usepackage{multirow}
\usepackage{balance}
\usepackage{threeparttable}
\usepackage{booktabs} 
\usepackage{epsfig}
\usepackage{arydshln}
\usepackage{url}
\usepackage{bbold}\usepackage{diagbox}
\input{insbox}

\usepackage{epstopdf}
\makeatletter
\newif\if@restonecol
\makeatother

\usepackage[linesnumbered, ruled, vlined]{algorithm2e}

\newcommand{\hide}[1]{} 
\newcommand{\vpara}[1]{\vspace{0.05in}\noindent \textbf{#1 }}

\newcommand{\secref}[1]{Section~\ref{#1}} 
\newcommand{\beq}[1]{\vspace{-0.03in}\begin{equation}#1\end{equation}\vspace{-0.03in}}
\newcommand{\beqn}[1]{\vspace{-0.04in}\begin{eqnarray}#1\end{eqnarray}\vspace{-0.04in}}

\newtheorem{problem}{Problem}
\newtheorem{definition}{Definition}

\newcommand{\PR}{CONNA$^r$(BP)}
\newcommand{\sPR}{CONNA$^r$(BP)\space}

\newcommand{\MPR}{CONNA$^r$(MFP)}
\newcommand{\sMPR}{CONNA$^r$(MFP)\space}

\newcommand{\MMR}{CONNA$^r$(MFMI)}
\newcommand{\sMMR}{CONNA$^r$(MFMI)\space}

\newcommand{\MMPR}{CONNA$^r$}
\newcommand{\sMMPR}{CONNA$^r$\space}

\newcommand{\MMPC}{CrossEntropy}
\newcommand{\sMMPC}{CrossEntropy\space}

\newcommand{\RC}{CONNA}
\newcommand{\sRC}{CONNA\space}

\newcommand{\JRC}{CONNA+Fine-tune}
\newcommand{\sJRC}{CONNA+Fine-tune\space}

\ifCLASSOPTIONcompsoc
\else
\fi
\ifCLASSINFOpdf
\else
\fi
\begin{document}
%
\title{CONNA: Addressing Name Disambiguation on The Fly}
%
%
%
%

\author{Bo~Chen\normalsize,
		Jing~Zhang*\normalsize,
		Jie~Tang\normalsize,\IEEEmembership{Senior Member,~IEEE},
		Lingfan Cai\normalsize, \\
		Zhaoyu Wang\normalsize,
		Shu Zhao\normalsize,
		Hong Chen\normalsize,
		Cuiping Li\normalsize

\IEEEcompsocitemizethanks{
\IEEEcompsocthanksitem Bo Chen, Jing Zhang, Lingfan Cai, Hong Chen and Cuiping Li are with Information School, Renmin University of China, Beijing, China. \protect\\
E-mail:\{bochen, zhang-jing, clfan, chong, licuiping\}@ruc.edu.cn\protect\\
*Corresponding author 
\IEEEcompsocthanksitem Jie Tang is with Department of Computer Science and Technology, Tsinghua University, and Tsinghua National Laboratory for Information Science and Technology (TNList), Beijing, China, 100084.\protect\\
E-mail:jietang@tsinghua.edu.cn
\IEEEcompsocthanksitem Zhaoyu Wang and Shu Zhao are with  School of Computer Science and Technology, Anhui University. \protect\\
E-mail:e17301101@stu.ahu.edu.cn, zhaoshuzs@ahu.edu.cn
}
\thanks{}}

%
%

\markboth{Journal of IEEE Transactions on Knowledge and Data Engineering,~Vol.~V, No.~N, January~2019}%
{Chen \MakeLowercase{\textit{et al.}}: CONNA: Addressing Name Disambiguation on The Fly}
%


\sloppy

\IEEEcompsoctitleabstractindextext{%
\begin{abstract}
Name disambiguation is a key and also a very tough problem in many online systems such as social search and academic search. Despite considerable research, a critical issue that has not been systematically studied is \textit{disambiguation on the fly} --- to complete the disambiguation in the real-time. This is very challenging, as the disambiguation algorithm must be accurate, efficient, and error tolerance. In this paper, we propose a novel framework --- CONNA --- to train a matching component and a decision component jointly via reinforcement learning. The matching component is responsible for finding the top matched candidate for the given paper, and the decision component is responsible for deciding on assigning the top matched person or creating a new person. The two components are intertwined and can be bootstrapped via jointly training. Empirically, we evaluate CONNA on two name disambiguation datasets. Experimental results show that the proposed framework can achieve a 1.21\%-19.84\% improvement on F1-score using joint training of the matching and the decision components. The proposed CONNA has been successfully deployed on AMiner --- a large online academic search system.

\end{abstract}

}

\maketitle

\IEEEdisplaynotcompsoctitleabstractindextext

%
\IEEEpeerreviewmaketitle

\section{Introduction}
\label{sec:intro}

Name disambiguation, aiming at disambiguating who is who, is one of the fundamental problems of the online academic network platforms such as Google Scholar, Microsoft Academic and AMiner. 
The problem has been extensively studied for decades~\cite{han2004two,huang2006efficient,louppe2016ethnicity,tang2012unified,wang2010constraint,wang2011adana,zhang2018name} and most of the works focus on how to group the papers belonging to same persons together into a cluster from scratch. However, online academic systems have already maintained a huge number of person profiles, which are made by the ``from scratch" algorithms or human beings. Out of the consideration of the computation and time cost of the real systems, it is not practical to re-compute the clusters from scratch for the new arriving papers every day. We need a more effective way to deal with the problem of name disambiguation on the fly. 

This paper takes AMiner as the basis to explain how we deal with the name ambiguity problem when continuously updating persons' profiles. AMiner is a free online academic search and mining system~\cite{tang2008arnetminer}, which has already extracted 133,204,120 researchers' profiles from the Web~\cite{tang2010combination} and integrated with 263,781,570 papers from heterogeneous publication databases~\cite{zhang2018name}. Currently, the newly arrived papers of AMiner are more than 500,000 per month. How to correctly assign these papers to the right persons in the system on the fly is a critical problem for many upper applications  such as expert finding, academic evaluation, reviewer recommendation and so on.  

\begin{figure}[t]
	\centering
	\includegraphics[width=0.5\textwidth]{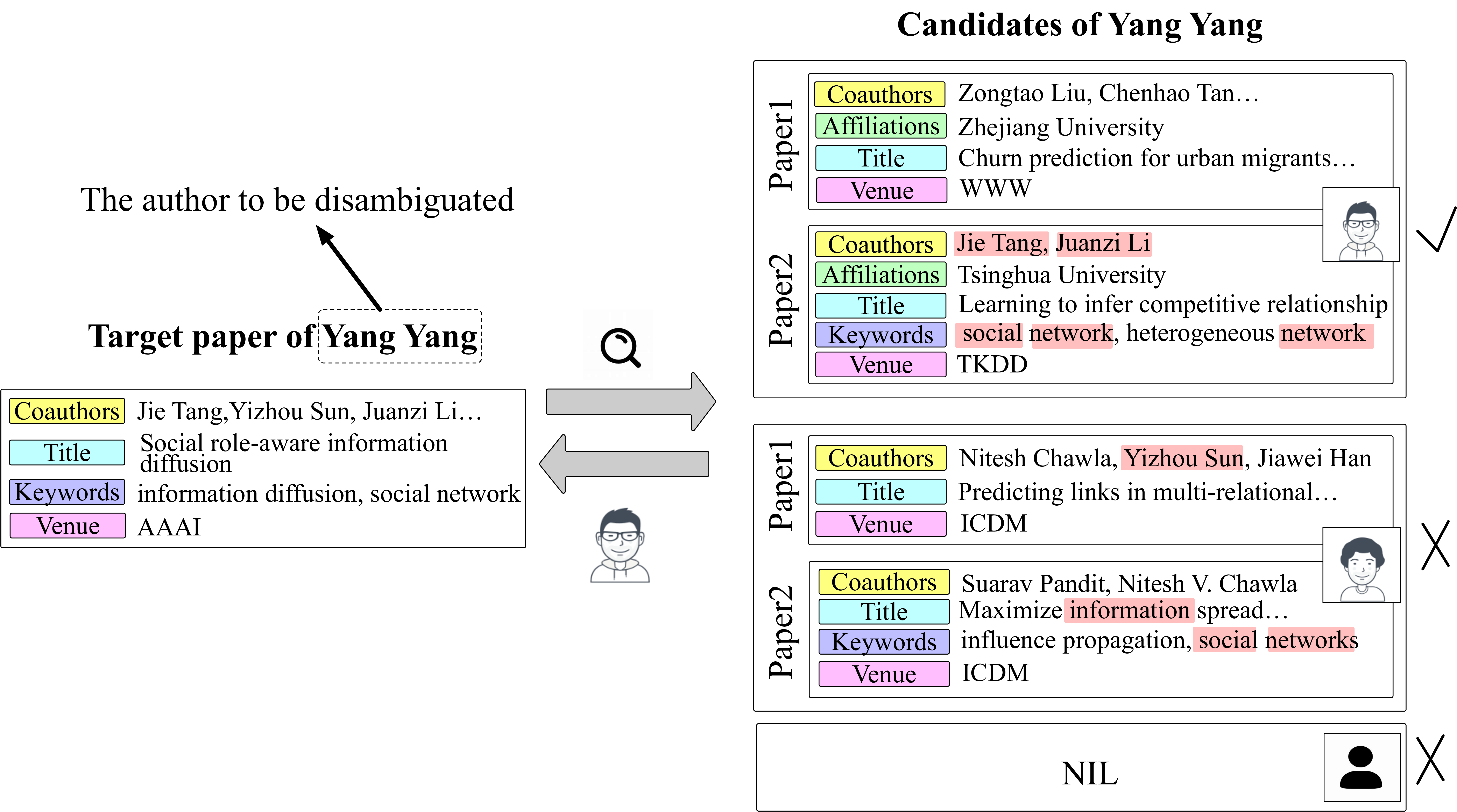}
	\caption{\label{fig:rankingchallenges} Disambiguation on the fly. Given a target
paper with target author as ``Yang Yang", we aim at searching for the right person
of ``Yang Yang" from the candidates, where the right person can be a
real person or a non-existing candidate denoted as NIL.}
\end{figure}

Existing methods on addressing the similar problem of anonymous author identification~\cite{chen2017task,zhang2018camel, zhao2019uncertainty} are possible solutions to continuously disambiguating papers on the fly.
However, they merely target at finding the top matched  person from all the candidates, but fail to deal with the situation when no right person exists, which is common in real academic systems. For example, the papers published by new researchers should not be assigned to any persons, as their profiles have not been established by the system. Thus, to assign a paper on the fly, we need to pay attention to not only find the top matched candidate, but also identify whether to assign the top matched candidate or create a new person. In other words, we  consider the absence of the right person from the candidates to be a distinct candidate, the so-called NIL candidate.
Figure~\ref{fig:rankingchallenges} illustrates the problem to be solved in the paper, where given a paper  with an author to be disambiguated, the returned right person can be a real person or a non-existing candidate denoted as NIL. Actually, in AMiner, in addition to the ``on-the-fly" assignment, we also perform a ``from scratch" algorithm to cluster ``NIL" papers into new profiles, and run an offline ``checking" algorithm to correct errors from historical profiles periodically. In general, AMiner performs a multi-strategy combining ``from scratch", ``on-the-fly" and ``checking" together to solve the complex continuous name disambiguation problem. In this paper, we only introduce the principle of ``on-the-fly" strategy under the assumption that the previously built profiles are correct, where the errors of the profiles are left to the ``checking" strategy.


To tackle the problem, we first investigate how to find the top matched candidate for a given target paper.
Straightforwardly, we can use the traditional feature-engineering methods to estimate the matching probability between each candidate and the target paper, and then return the top matched candidate. However, these methods are devoted to exactly matching the tokens between a paper and a person, which is too rigid and cannot handle the cases with similar semantics but different tokens.
The widely used representation-based models~\cite{chen2017task,zhang2018camel} can capture the soft/semantic matches through learning low-dimensional dense  embeddings, but they may contrarily hurt the performance of exact matching due to the highly compressed embeddings. For example in Figure~\ref{fig:rankingchallenges}, if only depending on the semantics of learned embeddings, we can infer that both of the candidates are interested in social network mining. However, it is apparent that the exact matches of the coauthor names or words, e.g., ``Jie Tang", ``Juanzi Li", ``social", ``network" between the target paper and the right person are more than those of the wrong person. Thus, a challenge is posed: \textit{how to capture both the exact matches and the soft matches in a principled way?} Simultaneously,  the effects of different fields are different. For example, the two matched coauthors in the right person make it significantly more confident than the wrong person with only one matched coauthor, compared with the matches in other fields. Besides, each person publishes multiple papers, which also take different effects. For example in Figure~\ref{fig:rankingchallenges}, in the papers of the right person, the effect of the second similar paper may be diluted by the first irrelevant one if combining all papers. Thus, \textit{an effective way to distinguish the effects of different fields of the attributes and different instances of the published papers is worth studying.} 

After obtaining the top matched candidate, we need to decide whether to assign the top matched candidate or NIL candidate to the target paper. 
The NIL problem is widely studied in entity linking, a similar problem that aims at linking the mentions extracted from the unstructured text to the right entities in a knowledge graph. We can adopt the similar idea to assign the NIL candidate to a  target paper if the score of the top matched person is smaller than a NIL threshold~\cite{gottipati2011linking,shen2013linking} or if the top matched person is predicted as NIL by an additional classifier~\cite{ratinov2011local}. Essentially, the first process of finding the top matched candidate tries to keep the relative distances between the right and the wrong persons of each target paper, and the later process of  assigning the top matched candidate or not devotes to optimize the absolute positions among top matched candidates of all target papers. Intuitively, the two processes can influence each other, and the errors of each process can be corrected by their interactions. However, \textit{none of the existing NIL solutions are aware of this and it is not clear how to correct the errors by the interactions between the two processes.}

To this end, in AMiner, we propose a joint model \sRC that consists of a matching component and a decision component to solve CONtinuous Name Ambiguity, i.e., name disambiguation on the fly, where ``on the fly" emphasizes the solved problem in the paper is different from name disambiguation ``from scratch". In the model, the matching component adopts an interaction-based deep learning model plus a kernel pooling strategy to capture both the exact and soft matches between a target paper and a candidate person and also  a multi-field multi-instance strategy to distinguish the effects of different attributes and different instances of papers. The decision component is trained on the similarity embeddings learned by the matching component, to further decide whether a top matched person is the right person or not. In addition, the errors of the proposed model can be self-corrected through  jointly fine-tune the two components by reinforcement learning. To summarize, the main contributions include:

\begin{itemize}

	\item We propose 
		\sRC consisting of a multi-field multi-instance interaction-based matching component and a decision component to address the problem of continuous name disambiguation. 
		With jointly fine-tuning of the two components by reinforcement learning, the errors of the two components can be self-corrected. 

	\item Experimental results on two large name disambiguation datasets show that \sRC compares favorably decision accuracy (+1.21\%-19.84\% in terms of F1) and matching accuracy (+ 3.80\%-49.90\% in terms of HR@1) against the baselines methods. CONNA is deployed on AMiner to assign papers on the fly now. All codes and data used in the paper are publicly available\footnote{https://github.com/BoChen-Daniel/TKDE-2019-CONNA}.
	

	
\end{itemize}

\section{Problem Formulation}
\label{sec:problem}

We introduce the definitions and the problem in this section.

\begin{definition}
	\textbf{Paper.} We denote a paper as $p$ associated with multiple fields of attributes, i.e., $p = \{A_1, \cdots, A_F\}$, where $A_f \in p$ represents the $f$-th attribute such as authors' names and affiliations, title, keywords, venue and so on. 
\end{definition}	
\begin{definition}
	\textbf{Target paper-author pair.} 	Given a paper $p$ with one of its authors denoted by $a$, we define a target paper-author pair as $\langle p,a \rangle$, where $p$ is the target paper and $a$ is the target author to be disambiguated. We abbreviate a target paper-author pair as a target pair henceforth.
\end{definition}
\begin{definition}
\textbf{Candidate Persons.}  Given a target pair $\langle p,a \rangle$, the corresponding candidate persons $C$ are those who are closely related to the target pair $\langle p,a \rangle$. Each candidate person $c_l \in C$ is composed of multiple papers, i.e., $c_l = \{p_1, \cdots, p_{n_l}\}$, where each paper $p_t = \{A_1, \cdots, A_F\}$ and $n_l$ is the number of papers published by $c_l$.
\end{definition}	

For a target pair $\langle p,a \rangle$, to find the right person from its candidate persons $C$, a straightforward way is to compare the coauthors' names of $a$ in $p$ with the coauthors' names of each candidate person in $C$\footnote{The names are treated as strings to be compared with each other.}.  The assumption is the more overlaps between the coauthors' names, the more likely the candidate is the right author of $p$. The similar idea is adopted in~\cite{liu2013s}, 
which found that if only using the users' names, 56\% same users with different accounts across the social networks can be correctly linked together. However, how can the names take effect in identifying the right person for the target pairs? 

\begin{figure}[t]
	\centering
	\includegraphics[width=0.5\textwidth]{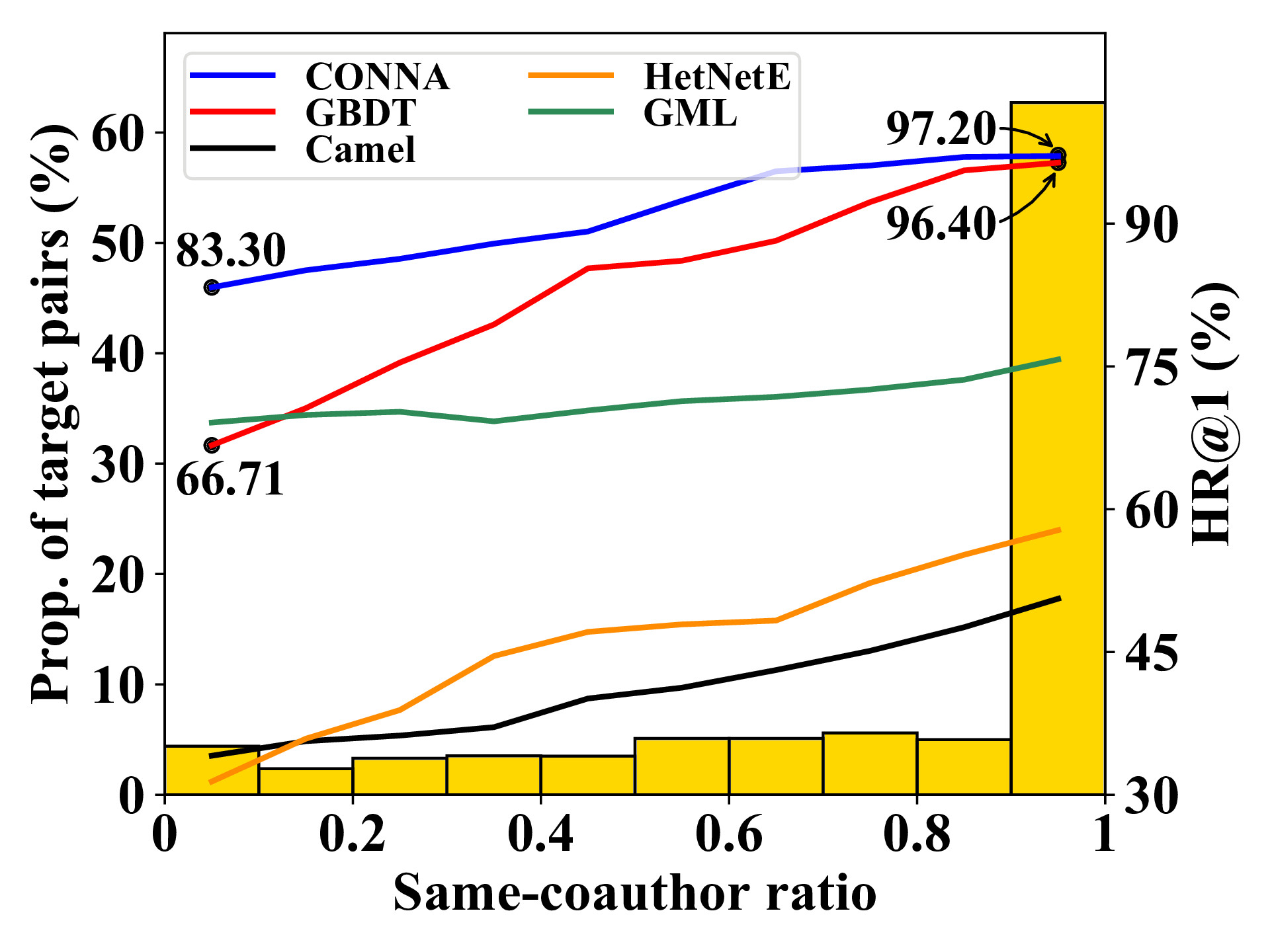}
	\caption{\label{fig:proportion_coauthor} Distribution of the same-coauthor ratio and the corresponding matching performance. {\small Yellow bar: Distribution of the same-coauthor ratio of the target pairs. Lines: HR@1 performances of different methods.}}
\end{figure}
To answer the question, we collect 100,000 target pairs from AMiner. For each target pair $\langle p,a \rangle$, we collect its candidate persons (Cf.~\secref{sec:overview} for candidate generation details) and calculate the same-coauthor ratio:

\beq{
	\label{eq:same-coauthor-ratio}
	\text{Same-coauthor ratio} = \frac{\max\limits_{c \in C} S_c  - \mathop{\text{second}}\limits_{c \in C} S_c}{\max\limits_{c \in C} S_c- \min\limits_{c \in C} S_c},
}

\noindent
where $S_c$ is the  number of the same coauthors of $a$ in $p$ with the candidate $c$. Same-coauthor ratio reflects the gap between the most similar candidate and the second similar candidate. The denominator is to normalize the gap calculated for different candidate lists into the same scale.
It will be easier to distinguish the right person from the other candidates when the same-coauthor ratio is larger.

Then we plot the distribution of the same-coauthor ratio for all the target pairs in Figure~\ref{fig:proportion_coauthor}, where X-axis indicates the same-coauthor ratio of a target pair, and Y-axis on the left denotes the proportion of the target pairs with a certain same-coauthor ratio. 
From the figure, we can see that although 62.72\% target pairs have large same-coauthor ratios, there are still 14.59\%  target pairs having small same-coauthor ratios. 
The coauthor-related features will hardly take effect when dealing with the target pairs with small same-coauthor ratios. For these target pairs, it is also not easy to leverage other features except the coauthor features.

To verify the above hypothesis, we estimate the probability of matching each candidate person to the target pair by GBDT based on several features such as the literal similarities between the title, venue, or the affiliations of the target pair and those of a candidate person besides the coauthor-related features, then evaluate whether the top matched candidate is the right person or not and show the evaluated metric, top 1 Hit Ratio (i.e., HR@1 on the right Y-axis) for different ranges of the same-coauthor ratio in Figure~\ref{fig:proportion_coauthor}. Clearly, we can see that the performance of GBDT decreases dramatically with the decrease of the same-coauthor ratio. The evaluated HR@1 is 66.71\% when the same-coauthor ratio is within (0, 0.1),  but is 96.40\% within (0.9,1.0). The results indicate that when the coauthors of the target pair and the right person are not similar, it is also difficult for feature-engineering methods to capture the similarities of other attributes. Thus, a more promising way to match each candidate with the target pair is required.

In addition to find the top matched candidate, we also need to consider the situation when no right person exists, which is usually ignored by existing author identification tasks~\cite{chen2017task,zhang2018camel}. Suppose an academic system establishes a profile for a researcher only if she/he has published at least one paper, a lot of papers written by the new researchers who publish papers for the first time, cannot be assigned to any existing person in the system. Thus, the right person should be either a real person or a non-existing person. In summary, the problem is defined as:

\begin{problem}
\textbf{Disambiguation on the fly.} Given a training set $\mathcal{D} = \{(\langle p,a \rangle,C)\}$, for each target paper-author pair $\langle p,a \rangle$ and the corresponding candidate persons $C$, the right person $c^*$ can be either a real person in $C$ denoted by $c^+$ or a non-existing person denoted by NIL, and other persons except $c^*$ in $C$ are the wrong persons denoted by $\{c^-\}$. The target is to learn a predictive function                                                                                                                                                                                                                                                                                                                                                                                                                        
\beq{
	\mathcal{F} : \{( \langle p, a \rangle, C)\} \rightarrow \{c^*\}
}to assign a target paper-author pair to its right person.
 \end{problem}

In our problem, $a$ is usually used to select candidate persons  and $p$ is used to extract features to match the candidates. 
To simplify the problem, we assume the historical papers assigned to the candidates are correct. However, historical errors cannot be avoided. Thus, we design an independent model  to check and correct the historical assignments repeatedly. The study is left in the future.

 \textsl{\section{\RC}}

In this section, we first give an overview of the end-to-end framework and then introduce the matching component which is to match the most possible candidate to the target pair and the decision component which is to decide  whether to assign the top matched candidate to the target pair or not respectively. Finally, we introduce how to self-correct the errors of the two components by jointly fine-tuning them through reinforcement learning. 
\subsection{Overview}
\label{sec:overview}
At first, given a target pair $\langle p,a\rangle$, the candidate persons $C$ are the persons having the relevant names with the target author $a$. We define the relevant names as simple variants of $a$'s name, including moving the last name to the first and keeping the initials of the names except for the last name. For example, the variants of ``Jing Zhang" include ``Zhang Jing", ``J Zhang" and ``Z Jing". For annotating a dataset for training and evaluating the models of name disambiguation, this simple candidate generation strategy can already result in enough challenging candidates. 

\begin{figure}[t]
	\centering
	\includegraphics[width=0.5\textwidth]{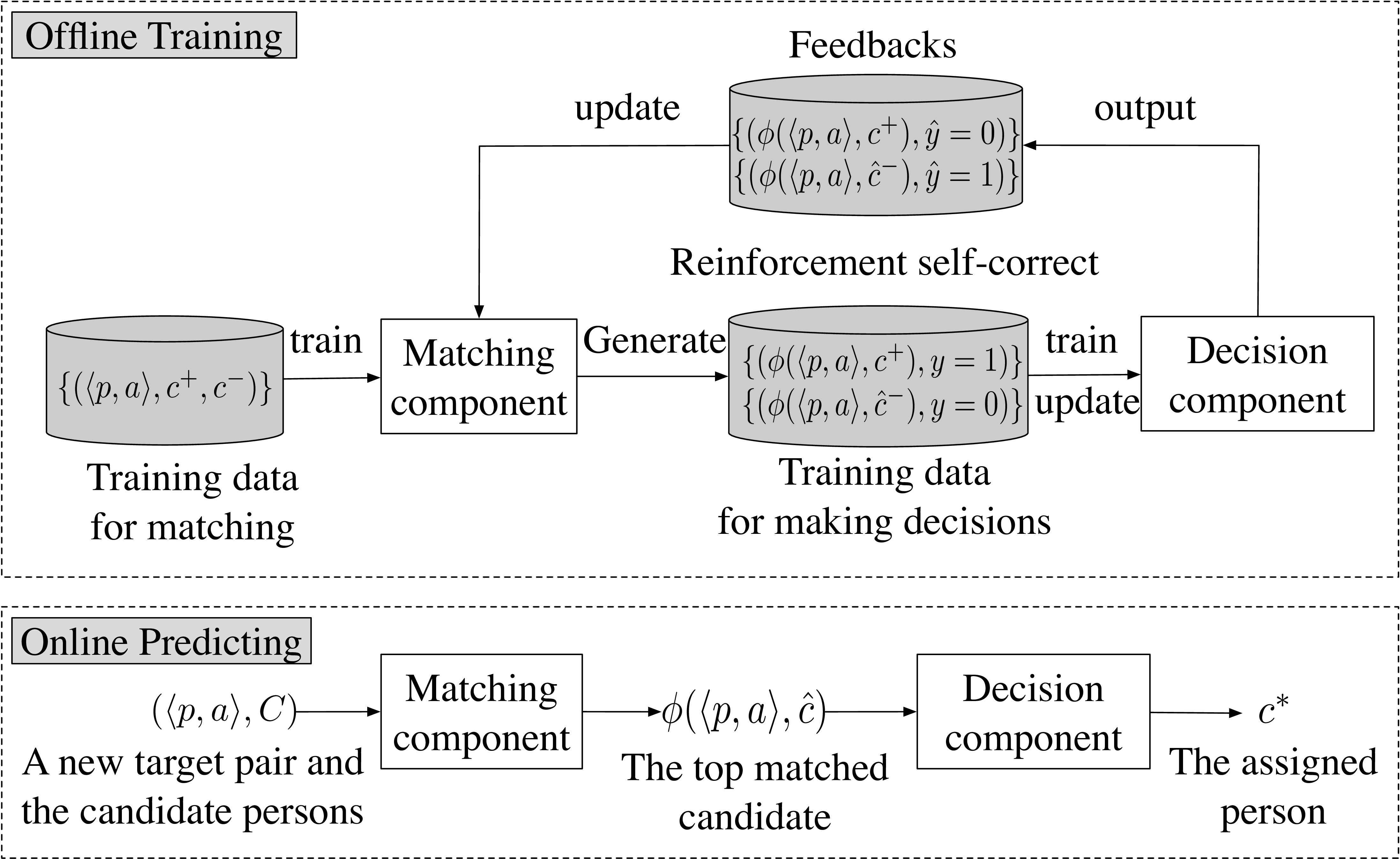}
	\caption{\label{fig:framework} The whole framework of training and predicting.}
\end{figure}

The whole process of name disambiguation is divided into offline training and online predicting, which is shown in Figure~\ref{fig:framework}. 
During the offline training process, we firstly train a matching component to estimate the probability of matching each candidate to the target pair and make the matching probability of the right person higher than those of the wrong persons for each target pair. The matching component constructs the training data from $\mathcal{D}=\{(\langle p,a \rangle,C)\}$ as a set of triplets $\mathcal{D}^r = \{(\langle p,a \rangle, c^+, c^-)\}$, where $\langle p,a \rangle$ is the target paper-author pair, $c^+$ is the real right person and $c^-$ is a wrong person from the candidates. The objective is to make $\langle p,a \rangle$ closer to $c^+$ than to $c^-$. 
Then, we train a decision component to accept  each sample $(\langle p,a \rangle,\hat{C}) \in \hat{\mathcal{D}}$  as the input and output a label $\hat{y}$ for the top matched person $\hat{c} \in \hat{C}$, where $\hat{C}$ is ranked by the trained matching component, $\hat{y}=1$ indicates $\hat{c}$ is the right person and $\hat{y}=0$ indicates $\hat{c}$ is the wrong person. 
We construct the  training data  $D^c$ for the decision component by extracting $(\langle p,a \rangle,c^+)$ as the positive instance (i.e., $y=1$) and $(\langle p,a \rangle,\hat{c}^-)$ as the negative instance (i.e., $y=0$) from each sample $(\langle p,a \rangle,\hat{C})$, where  $\hat{c}^-$ indicates the top matched wrong person in $C$. 
Finally, we fine-tune the matching component based on the feedback (i.e., error cases) of the decision component, and then fine-tune the decision component based on the updated output of the matching component. Essentially, the matching component tries to keep the relative distances between the right and the wrong persons of each target pair, and the decision component devotes to optimize the absolute positions between the top matched persons of all the target pairs found by the matching component. 

During the online predicting process, to disambiguate a target pair $\langle p,a \rangle$, the matching component firstly finds out the top matched candidate person $\hat{c}$, then based on the similarity features $\phi(\langle p,a \rangle,\hat{c})$ output by the matching component, the decision component will predict the label $\hat{y}$ for $\hat{c}$ and finally assign the person $c^*$ to $\langle p,a \rangle$, where $c^* = \hat{c}$ if $\hat{y}=1$ and $c^* = $ NIL otherwise.

\subsection{Matching}

\vpara{Basic Profile Model (BP).}
\label{sec:interaction}
Let's imagine how humans assign a paper to a person. The humans usually browse all the papers published by the person to understand her/his affiliation, overall research interest, and frequently collaborated authors, then comparing them with those of the paper. In other words, humans directly compare the person's profile with the target pair, which can guide us to build our model. Thus, we name the model as the basic profile model. Specifically, we merge all the attributes of a paper and divide them into a set of tokens to represent the paper, and then merge the tokens of all the papers of a person into a unified set of tokens to represent the person's profile. Based on the token-based representations of the target paper and the person, we can estimate the similarity between them. 
Note a complete author name or a word in titles, keywords, venues and affiliations is viewed as a token.

Some metrics such as Jaccards Coefficient~\cite{Salto:1983} and cosine similarity~\cite{Salto:1983} can easily capture the exact matches. However, they suffer from the sparsity of the token-based representations. For example, the similarity is zero if two representations do not contain any same tokens, even if they are semantically similar. On the other hand, recently, some representation-based models~\cite{hu2014convolutional,huang2013learning} can successfully capture the soft/semantic similarities, as they embed the high-dimensional sparse features into low-dimensional dense representations. Through training on the labeled data, the model can reduce the distance between the semantically similar inputs in the low-dimensional space. However, these models may suffer from the problem of semantic drift. For example, two token-based representations with many overlapped tokens may become dissimilar after being embedded by the model, as the global representation may dilute the  effect of the exact same tokens by other different tokens. In summary, the above two types of methods are good at either exact matching or soft matching. To capture both the exact and soft matches, we adopt the interaction-based  models~\cite{dai2018convolutional,hu2014convolutional,xiong2017end} widely used in information retrieval. The interaction-based models first build a similarity matrix between each candidate person and the target pair and then apply an aggregation function to extract features from the matrix. These models avoid learning the global representations,  thus can reduce the issue of semantic drift. 

\textit{Similarity Matrix.}
We represent the matches between each candidate and the target pair as a similarity matrix $\mathbf{S}$, with each element $\mathbf{S}_{ij}$ standing for the basic interaction, i.e., the cosine similarity 
$\mathbf{S}_{ij} = \frac{\mathbf{p}_i \cdot \mathbf{c}_j }{||\mathbf{p}_i||\cdot ||\mathbf{c}_i||}$ between $\mathbf{p}_i$ and $\mathbf{c}_j$, where $\mathbf{p}_i$ represents the embedding of the $i$-th token in the target pair $\langle p,a \rangle$ and $\mathbf{c}_j$ represents the embedding of the $j$-th token in the candidate person $c$, which can be pre-trained by Word2Vec~\cite{mikolov2013efficient,li2019scaling} or BERT~\cite{devlin2019bert}.

\begin{figure}[t]
	\centering
	\includegraphics[width=0.47\textwidth]{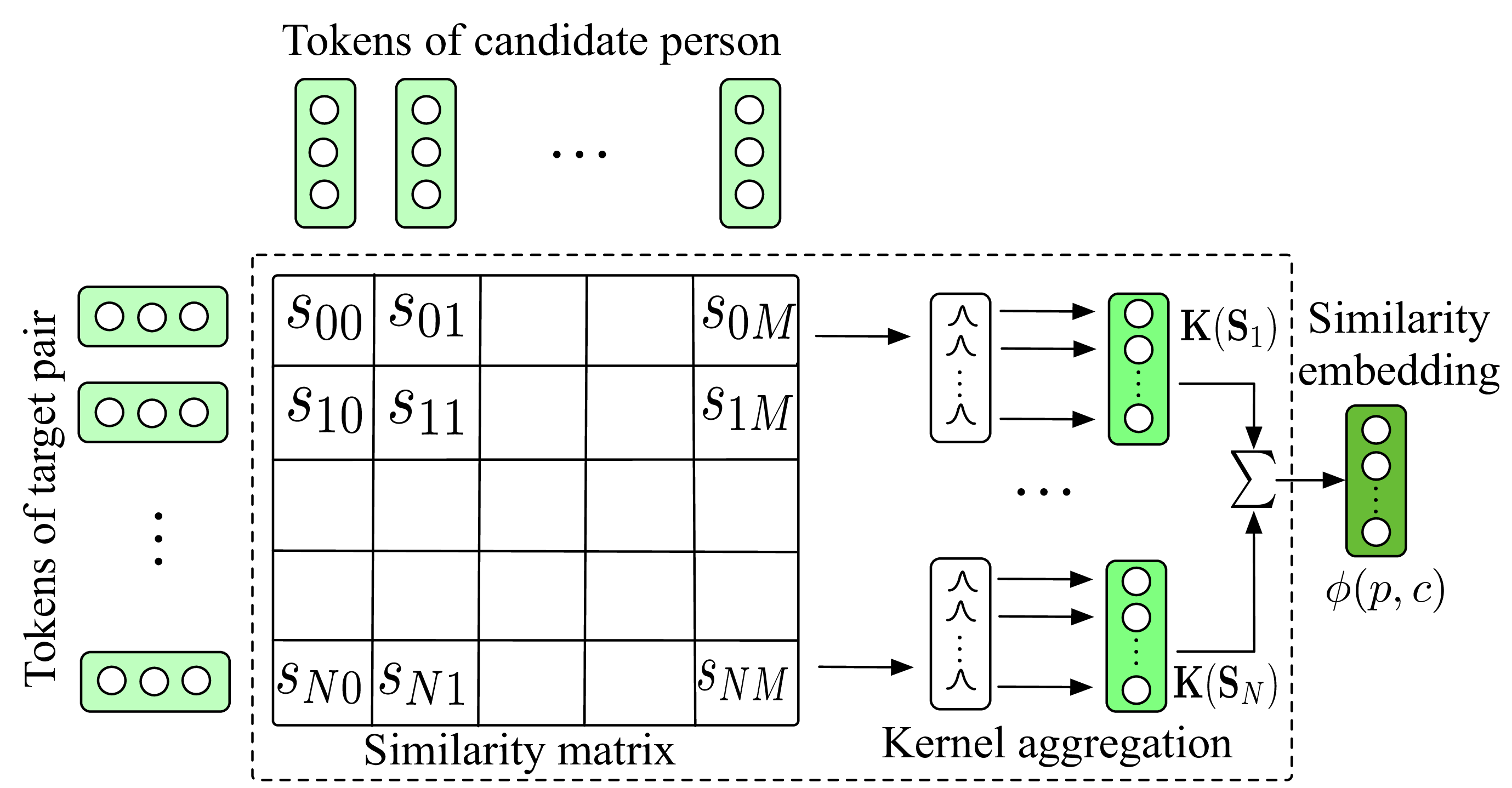}
	\caption{\label{fig:profilemodel} The basic profile model.}
\end{figure}

\textit{Aggregation Function.}
For sentence matching, CNN~\cite{hu2014convolutional,pang2016text} and RNN~\cite{wan2016match} are widely used as aggregation functions to extract matching patterns from the similarity matrix. However, different from sentence matching, title, keywords, venue and affiliation are all short text. We need to pay more attention to the occurrence of the exact same or semantically similar tokens. Thus we adopt an RBF kernel aggregation function~\cite{xiong2017end} to extract features. Specifically, the $i$-th row $\mathbf{S}_i = \{S_{i0},\cdots, S_{iM}\}$ of the similarity matrix --- the similarities between the $i$-th token of the target pair and each token of the candidate person, is transformed into a feature vector $\mathbf{K}(\mathbf{S}_i)$, with each of the $k$-th element $K_k(\mathbf{S}_i)$ being converted by the $k$-th RBF kernel with mean $\mu_k$ and  variance $\sigma_k$. Then the feature vectors of all the tokens in the target pair are summed up into the final similarity embedding $\phi(\langle p,a \rangle,c)$, i.e., 

\beqn{
	\phi(\langle p,a \rangle,c) &=& \sum_{i=1}^{N} \log \mathbf{K}(\mathbf{S}_i) ,\\ \label{eq:phi}
	\mathbf{K}(\mathbf{S}_i)&=& \{K_1(\mathbf{S}_i), \cdots, K_K(\mathbf{S}_i)\}, \\ \label{eq:K_vector}
	K_k(\mathbf{S}_i) &=& \sum_{j=1}^{M} \exp \left[ - \frac{(S_{ij} - \mu_k)^2}{2\sigma_k^2}\right]. \label{eq:K}
}

The kernel with $\mu=1$ and $\sigma \rightarrow 0$ only considers the exact matches between tokens, and others, e.g., with $\mu=0.5$, counts the number of tokens in the candidate person whose similarities to a queried token in the target paper are close to 0.5.  Thus, the kernel aggregation not only emphasizes the effect of exact matching but also captures the soft matches.  Figure~\ref{fig:profilemodel} illustrates the model.

\begin{figure}[t]
 	\centering
 	\includegraphics[width=0.47\textwidth]{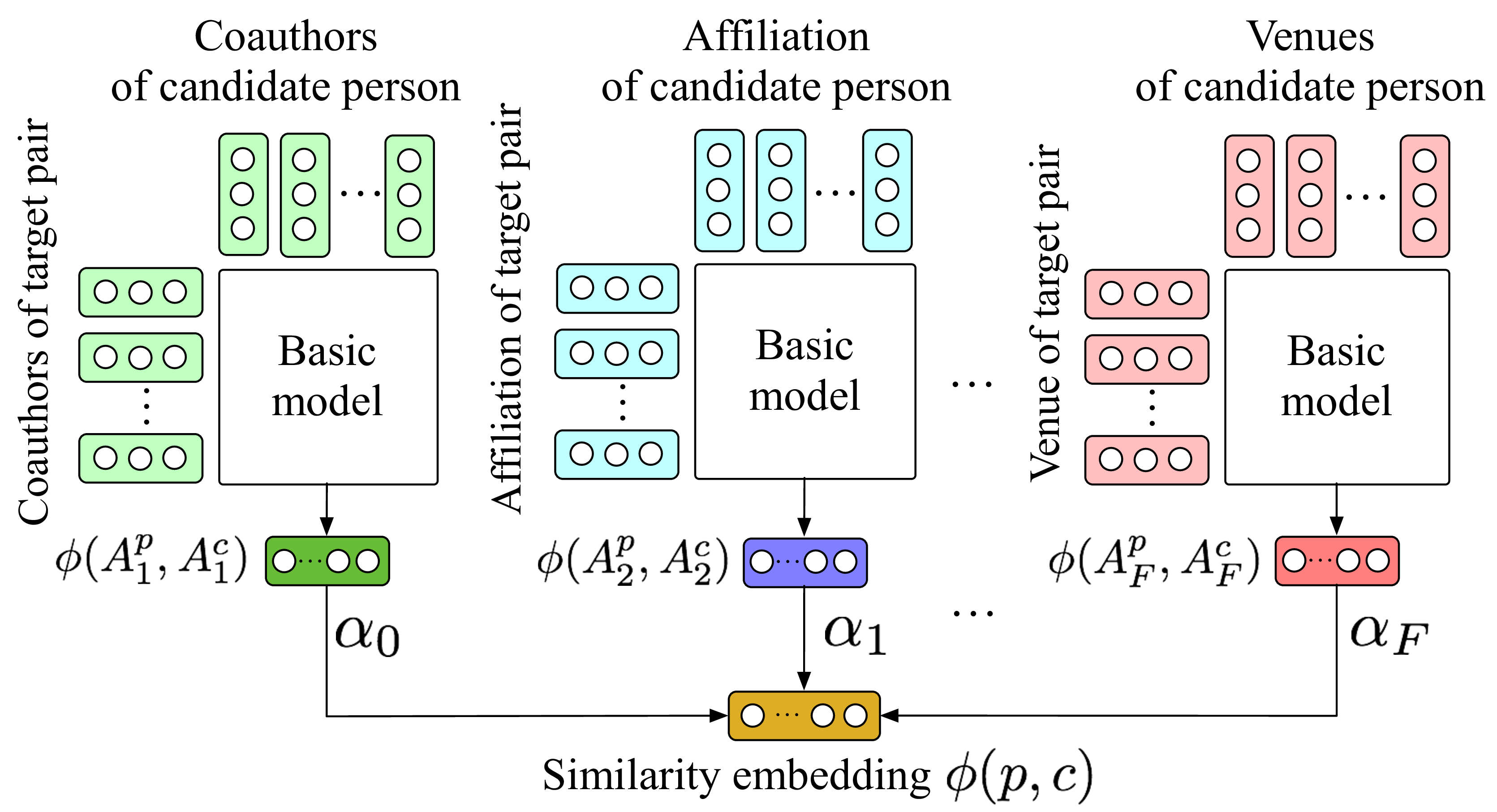}
 	\caption{\label{fig:multi-field} The multi-field profile model.}
 \end{figure}

\vpara{Multi-field Profile Model (MFP).}
\label{sec:multi-field}
The basic profile model does not distinguish different fields of attributes but groups them together. However, it is not necessary to compare different attributes, such as comparing authors with venues. Moreover, it takes more effect to compare coauthor names than other attributes. So we build a basic profile model on each field of the attributes respectively, i.e. different attributes are not allowed to be compared, then aggregate the similarity embeddings together by the corresponding attention coefficients estimated by:

\beqn{\label{eq:attention}
	  \alpha_f &=& \frac{ \exp (w\phi(A_f^p, A_f^c) + b)}{ \sum_{f}\exp(w\phi(A_f^p,A_f^c) + b)}, \\ \nonumber
	  \phi(\langle p,a \rangle,  c) &=& \sum_f \alpha_f \phi(A_f^p, A_f^c), \nonumber
}

\noindent where $\phi(A_f^p, A_f^c)$ denotes the similarity embedding between $A_f^p$ and $A_f^c$ with $A_f^p$ being the $f$-th field of $p$ and $A_f^c$ being that of the candidate person $c$. Notations $w$ and $b$ denote the parameters. The model is named as multi-field profile model and is shown in Figure~\ref{fig:multi-field}.

\begin{figure}[t]
	\centering
	\includegraphics[width=0.47\textwidth]{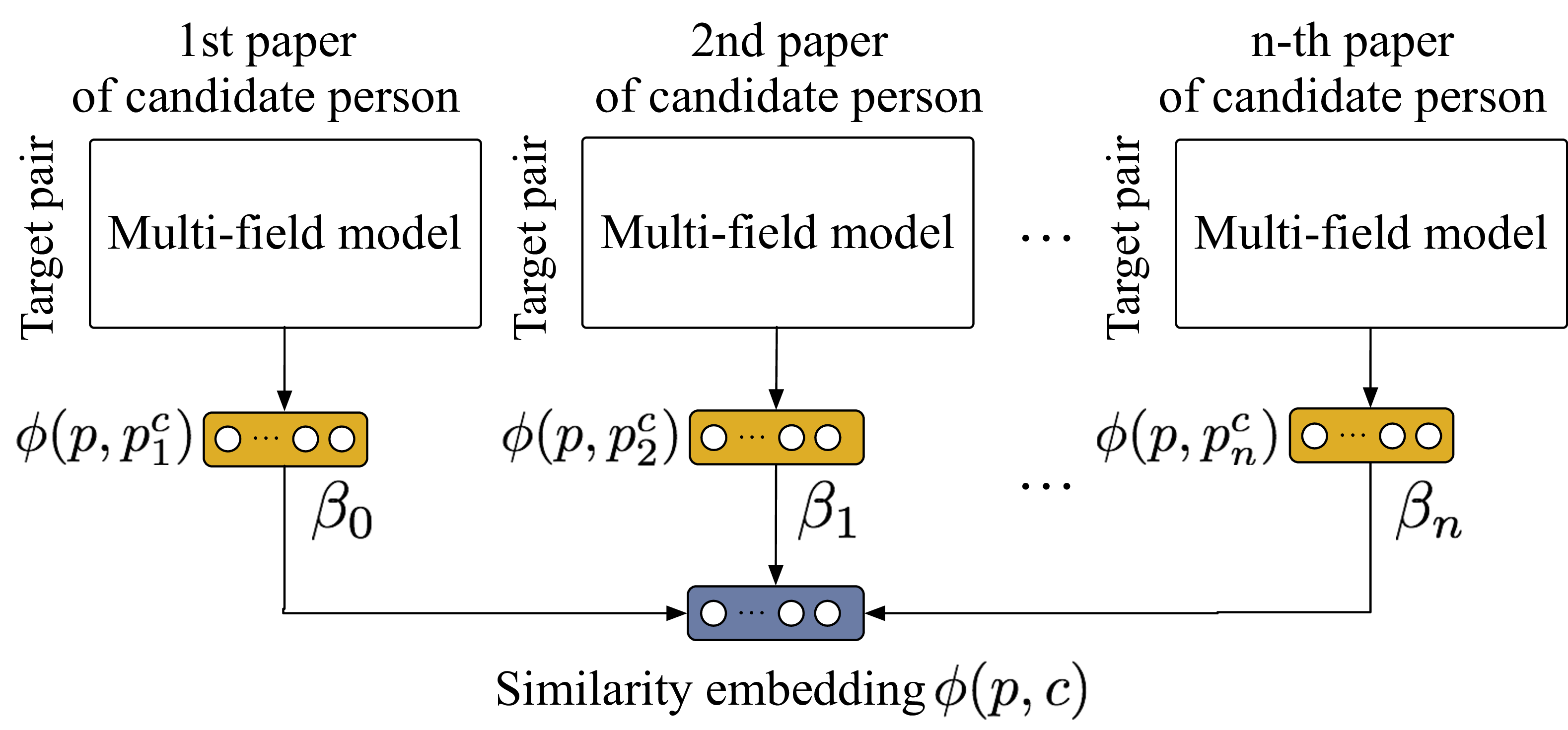}
	\caption{\label{fig:multi-instance} The multi-field  multi-instance model.}
\end{figure}

\vpara{Multi-field Multi-instance Model (MFMI).}
\label{sec:multi-field-multi-instance}
A person usually publishes multiple papers. Some persons even publish papers of multiple topics on multiple fields of venues and collaborate with multiple communities of persons. In this scenario, a target paper can be only similar to a small part of a person's diverse profile, but is totally irrelevant to other parts of the profile. However, the multi-field profile model may dilute the similarity with this small part  when summing the similarities with all the tokens in a person's profile together by Eq.\eqref{eq:K}. To reduce the impact from the irrelevant papers, we build a multi-field model between the target pair and each published paper of the candidate person, and then aggregate the resultant similarity embeddings of all the published papers by their corresponding attention coefficients, which are estimated the same as Eq.\eqref{eq:attention}. The model is named as the multi-field multi-instance model and is shown in Figure~\ref{fig:multi-instance}. 


\vpara{The Combination Model (CONNA$^r$).}
\label{sec:combinationmodel}
Essentially, the multi-field profile model captures the global similarities between the
target pair and a person's profile, while the multi-field multi-instance model considers the local similarities between the target pair and each of the papers published by a person. Both of them can be explained
intuitively, thus we can combine their output similarity
embeddings together as the final feature embedding.
We summarize different component variants in Table \ref{tb:modelcomponents}.

\vpara{Loss Function.}
\label{sec:triplets}
We use the triplet loss function to optimize the matching component. Similar ideas has been also used in~\cite{chen2017task,zhang2018camel,zhang2018name}. 
Let $\mathcal{D}^r$ be a set of triplets with each triplet denoted as $(\langle p,a \rangle, c^{+}, c^{-})$, where $c^+$ is the right person of the target pair $\langle p,a \rangle$ and $c^-$ is a wrong person sampled from the candidates, the triplet loss function $\mathcal{L}(\Theta)$ is defined as:

{\footnotesize \beqn{
	\label{eq:rankloss}
	\mathcal{L}(\Theta) \!
	&=& \!\!\!\!\!\!\!\!\!\!\!\sum_{(\langle p,a \rangle,c^+,c^-) \in \mathcal{D}^r} \mathcal{L}_{\Theta}(\langle p,a \rangle,c^+,c^-)  \\
	&=& \!\!\!\!\!\!\!\!\!\!\!\!\!\!\!\!\sum_{(\langle p,a \rangle,c^+,c^-) \in \mathcal{D}^r} \!\!\!\!\!\!\!\!\!\max \{ 0, g(\phi(\langle p,a \rangle, c^{-}))-g(\phi(\langle p,a \rangle, c^{+}))+m)\},	\nonumber
}}

\noindent where $g$ is defined to be a non-linear function to transform the similarity embedding $\phi$ into a one-dimension matching score that can be compared between the positive pair $(\langle p,a \rangle, c^{+})$ and the negative pair $(\langle p,a \rangle, c^{-})$. Notation $\Theta$ indicates the parameter of the matching component and $m>0$ is a margin enforcing a distance between positive pairs and negative pairs. We optimize the triplet loss instead of directly optimizing the cross-entropy loss between the output matching score and the true label, as we aim at finding the top matched candidate from all the candidates for each target pair, thus the objective should be keeping a relative order within the candidate persons of each target pair instead of keeping a global order among all the $(p,c)$ pairs. The triplet loss is more direct and close to our objective than the cross-entropy loss.

\subsection{Decision}
\label{sec:classification}

The decision component is built upon the output of the matching component to identify the right person, who can be either the top matched real person or NIL. 
The candidate persons $C$ of each sample $(\langle p,a \rangle,C) \in \mathcal{D}$ are ranked into $\hat{C}$ based on the matching probabilities estimated by the matching component. Note for the samples with $c^* = c^+$,  the real right person $c^+$ may be ranked the first or not. Then the decision component is trained to predict the first ranked person $\hat{c} \in \hat{C}$ to be a right person (i.e., $\hat{y}=1$) or a wrong person (i.e., $\hat{y}=0$). To achieve the goal, we construct the training data $\mathcal{D}^c$ from the ranked dataset $\hat{\mathcal{D}}=\{(\langle p,a \rangle,\hat{C}) \}$. Specifically, from each sample $(\langle p,a \rangle,\hat{C})$, we extract $(\langle p,a \rangle,c^+)$ as the positive instance (i.e., $y=1$) and extract $(\langle p,a \rangle,\hat{c}^-)$ as the negative instance (i.e., $y=0$), where $\hat{c}^-$ indicates the first ranked wrong person in $C$. 
In another words, the positive instances are only extracted from the samples with $c^* = c^+$, while the negative instances are extracted from both the samples with $c^* = c^+$ and the samples with $c^* = \text{NIL}$. 
For an instance $(\langle p,a \rangle,c)$, we use the similarity embedding $\phi(\langle p,a \rangle,c)$ output by the matching component as its feature. Thus,  $\mathcal{D}^c=\{(\phi(\langle p,a \rangle,c^+),y=1)\} \cup \{(\phi(\langle p,a \rangle,\hat{c}^-),y=0)\}$.  Then we train a multi-layer perceptron $h(\Phi)$:

\beq{
	\label{eq:decision}
		h(\Phi): \{\phi(\langle p,a \rangle,c)\} \rightarrow \{y\},
}

\noindent where $y$ is the label of the instance $(\langle p,a \rangle,c)$, whose value equals 1 if $(\langle p,a \rangle,c)$ is a positive instance and 0 otherwise.

\begin{table}
	\newcolumntype{?}{!{\vrule width 1pt}}
	\newcolumntype{C}{>{\centering\arraybackslash}p{3.5em}}
	\caption{
		\label{tb:modelcomponents} Matching component  variants of \RC.
		\normalsize
	}
	\centering  \scriptsize
	\renewcommand\arraystretch{1.0}
	\begin{tabular}{ll}
		\toprule
		Component variants	& Key idea\\
		\midrule
		Basic Profile (BP)
		& The basic interaction-based model
		\\ 
		Multi-field Profile (MFP)
		& Build BP for each field
		\\
		Multi-field Multi-instance (MFMI)
		& Build MFP for each instance
		\\
		CONNA$^r$
		& Combine MFP and MFMI 
		\\
		\bottomrule
		
	\end{tabular}
	
\end{table}

\subsection{Reinforcement Self-correction}
\label{sec:joint}
We finally fine-tune the two components by jointly training them to correct their errors by themselves.
The matching component can be viewed as  the generator to generate the ranking list. Without the decision component, the triplet loss in Eq.\eqref{eq:rankloss} is used to measure whether the ranking list is good or not. However, as the final objective is to determine whether the top ranked candidate is the right person or not, the triplet loss is not enough to verify the effect. Fortunately, we can use the prediction result of the top ranked candidate by the decision component as the delayed feedback to the ranking results of the matching component. 
Specifically, we can punish the ranking list with the wrongly predicted top candidate and reward the ranking list with the correctly predicted top candidate. Then based on the reward we update the matching component, expecting the ranking lists generated by the matching component to the decision component are more accurate. Followed by the motivation, we propose fine-tuning the two components via reinforcement learning. Specifically, the objective is to maximize the expected reward of the ranking lists generated by the matching component:

\beq{
	\label{eq:expectation}
	J(\Theta) = \sum_{(\langle p,a \rangle,\hat{C})) \in \hat{\mathcal{D}}} p_{\Theta}(\langle p,a \rangle,\hat{C})R(y, \hat{y}),
}

\noindent where $\hat{\mathcal{D}}$ is the ranked training data, $p_{\Theta}(\langle p,a \rangle,\hat{C})$ is the probability of generating the ranking list $\hat{C}$ of the target pair $\langle p,a \rangle$ by the matching component, and $R(y,\hat{y})$ is the reward function defined as follows:

\beq{
	\label{eq:reward}
	R(y, \hat{y}) =  \left\{\begin{array}{cl} 
			1 & \hat{y} = y;   \\
			0 & \mbox{otherwise.}   
		\end{array}\right.
}

\noindent where $\hat{y}$ is the predicted label for the top-ranked candidate $\hat{c}$ of $\hat{C}$ and $y$ is the ground truth label. The defined reward function encourages the matching component to float the right person at the top and push the wrong person away from the top. The policy gradient algorithm~\cite{sutton:00} is adopted to optimize the expected reward in Eq.\eqref{eq:expectation}, whose gradient is calculated as:

\beqn{
	\label{eq:gradient} \nonumber
	\nabla_{\Theta} J(\Theta) \!\!\!\!&=& \!\!\!\!\!\!\!\!\!\!\!\sum_{(\langle p,a \rangle,\hat{C}) \in \hat{\mathcal{D}}} R(y, \hat{y})\nabla p_{\Theta}(\langle p,a \rangle,\hat{C}), \\ 
	\!\!\!\!&\simeq&\!\!\!\!\!\!\!\!\!\!\!\!\!\!\!\! \sum_{(\langle p,a \rangle,\hat{c}, c^-) \in \mathcal{D}^r} \!\!\!\!\!\!\!R(y, \hat{y})\nabla \mathcal{L}_{\Theta}(\langle p,a \rangle,\hat{c}, c^- ).
}

Since the probability of a ranking list $\hat{C}$ is not easy to be estimated, we transform $\hat{C}$ into a set of triplets, with each triplet including the target pair $\langle p,a \rangle$, the top ranked candidate $\hat{c} \in \hat{C}$ and a negative candidate $c^{-} \in \hat{C}$. Then the loss of a triplet in Eq.\eqref{eq:rankloss} is calculated and the losses of all the triplets are summed up to approximately measure the ranking performance of $\hat{C}$. Thus, the gradient $\nabla p_{\Theta}(\langle p,a \rangle,\hat{C})$ is approximated by $ \nabla \mathcal{L}(\langle p,a \rangle,\hat{c}, c^-)$ of all the triplets  in $\hat{C}$.
Then the parameters $\Theta$ of the matching component can be updated by the gradient. After the matching component is tuned, the decision component is also updated based on the updated similarity embeddings output by the matching component. Algorithm~\ref{algo:joint_model} illustrates the joint training process, where we firstly pre-train the matching component and the decision component, and then jointly fine-tune the two components together.


\begin{algorithm}[t]
	{\small \caption{Reinforcement Joint Training\label{algo:joint_model}}
		\KwIn{A training set $\mathcal{D} = \{(\langle p,a \rangle,C)\}$.}
		\KwOut{A matching component and a decision component parametrized by $\Theta$ and $\Phi$ respectively.}
		Build $\mathcal{D}^r=\{(\langle p,a \rangle,c^+,c^-)\}$ from $\mathcal{D}$;\\
		Pre-train $\Theta$ of the matching component on $\mathcal{D}^r$;\\
		Rank $\mathcal{D}$ by the matching component to generate  $\hat{\mathcal{D}}$;\\
		Build $\mathcal{D}^c = \{(\phi(\langle p,a \rangle,c),y)\}$ from $\hat{\mathcal{D}}$;\\
		Pre-train $\Phi$ of the decision component on $\mathcal{D}^c$;\\
		\Repeat{Convergence}{
		\For{ $(\langle p,a \rangle,\hat{C})\in \hat{\mathcal{D}}$ }{
				Predict $\hat{y}$ for $\hat{c}$ by the decision component;\\
				Calculate $R(y, \hat{y})$ by Eq.\eqref{eq:reward};\\
				Calculate $\nabla_{\Theta} J(\Theta)$ by Eq.\eqref{eq:gradient};\\
				$\Theta \rightarrow \Theta + \mu \nabla_{\Theta} J(\Theta)$, where $\mu$ is the learning rate;\\
			}
			Re-rank $\mathcal{D}$ to generate $\hat{\mathcal{D}}$ by the matching component;\\
			Re-generate $\mathcal{D}^c$ from $\hat{\mathcal{D}}$;\\
			Update $\Phi$ of the decision component on $\mathcal{D}^c$;\\
		}
		}
	\end{algorithm}

\section{Experiment}
\label{sec:exp}

\hide{
\begin{figure}[t]
	\centering
	\includegraphics[width=0.47\textwidth]{Figures/reinforce}
	\caption{\label{fig:reinforce} An Illustration of Reinforcement Joint Fine-tuning.}
\end{figure}
}

All codes and data used in the paper are publicly available\footnote{https://github.com/BoChen-Daniel/TKDE-2019-CONNA}.

\subsection{Experimental Settings}
\subsubsection{Datasets} 
We evaluate CONNA on two name disambiguation datasets:

\textit{OAG-WhoIsWho}\footnote{https://www.aminer.cn/whoiswho}: Is the largest human-annotated name disambiguation dataset so far, which is consist of 608,363 papers belonging to 57,138 persons of 642 common names. Existing work either leverage the disambiguating results by algorithms in some well-known academic websites such as Scopus~\cite{reijnhoudt2014}， CiteSeerX~\cite{zhang2017name}, Web of Science~\cite{Backes:2018} and PubMed~\cite{torvik2009author}, or annotate a much smaller datasets by human beings, such as 8,453~\cite{han2004two}, 6,921~\cite{kang2011construction}, 7,528~\cite{tang2012unified} and 2,946 annotated persons~\cite{muller2017data}. Compared to the most popular KDD Cup 2013 challenge dataset, the OAG-WhoIsWho is also superior to it both in quantity (608,363 vs 424,384 in terms of the number of papers) and quality (fully human-labeled vs partially human-labeled). We annotate the dataset as follows.
From the AMiner system, we choose 642 highly ambiguous names, create the relevant names by the candidate generation strategy in \secref{sec:overview} and select all the authors for each name, collect all the papers assigned for each author and extract title, authors, organizations, keywords and abstract for each paper. We also collect all the unassigned papers for each name from AMiner. Since the assigned papers may be wrongly assigned and the papers are not fully assigned, additional efforts are needed to clean and reassign the papers. First, we clean the dataset by removing the wrongly assigned papers or splitting the papers of an author into different clusters. Second, we annotate the unassigned papers or merge the papers of two authors. We aim to clean the dataset as much as possible but increase the highly reliable assignments. According to the purpose, we only hire one annotator to perform the cleaning step, but hire three annotators to perform the assignment step respectively and then obtain the final results by majority voting their annotations. Besides, an annotation tool is developed to recommend highly reliable removing, splitting, assigning or merging operations to the annotators to simplify the human annotation process\footnote{https://www.aminer.cn/annotation}.

\begin{figure}
	\centering
	\subfigure[]{\label{subfig:kdd-cup}
		\includegraphics[width=0.22\textwidth]{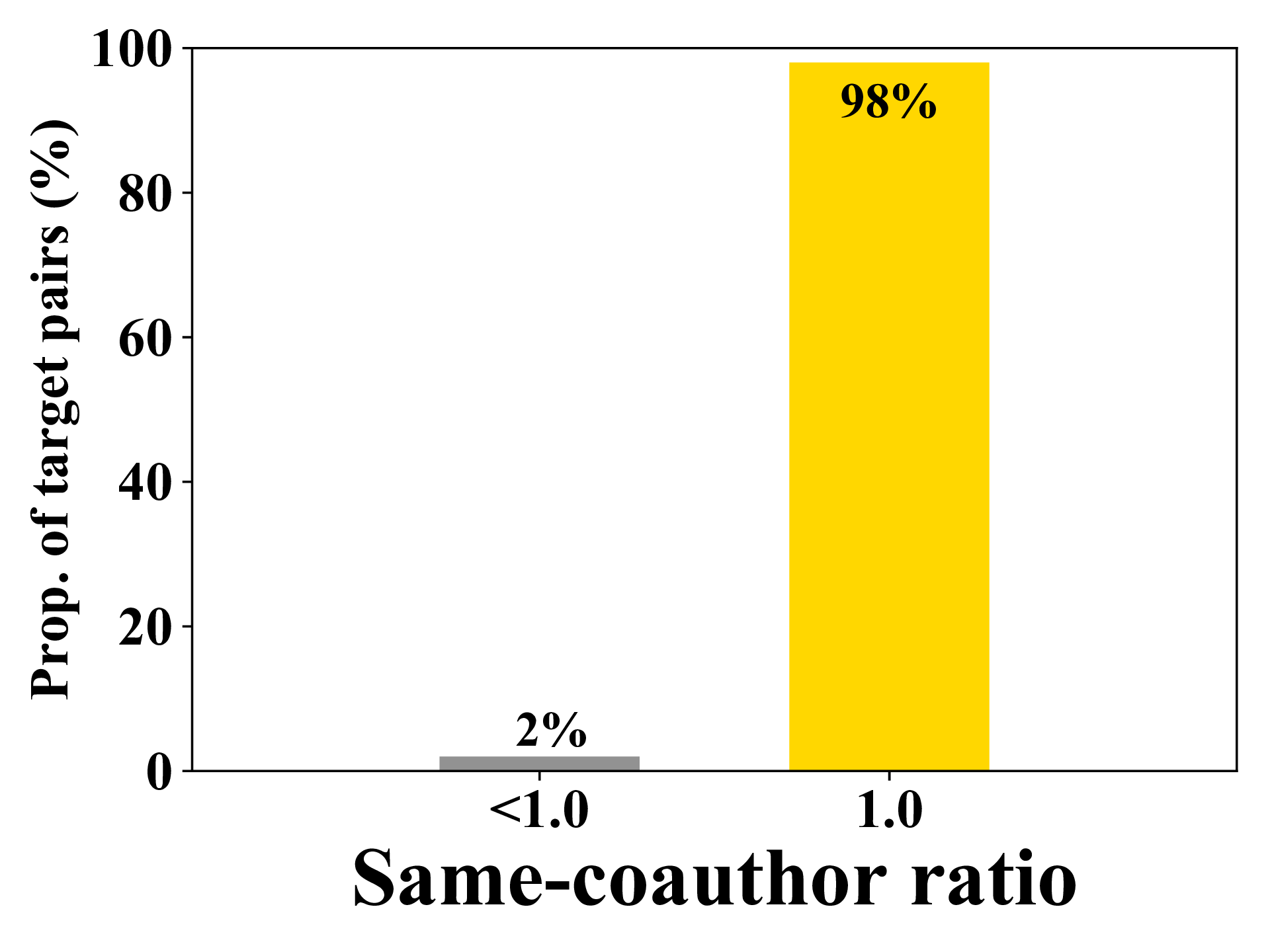}
	}
	\centering
	\subfigure[]{\label{subfig:name_attr}
		\includegraphics[width=0.22\textwidth]{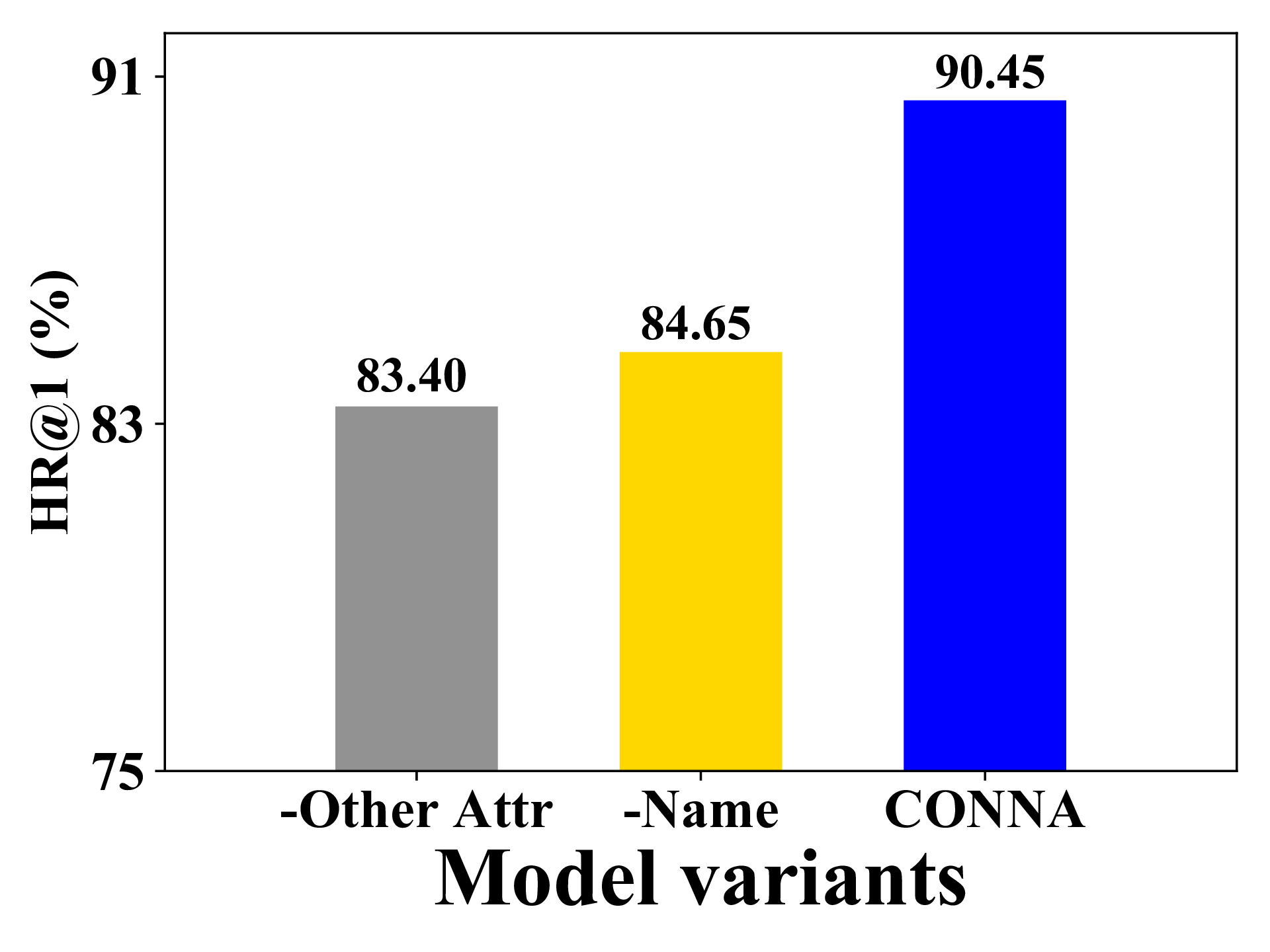}
	}
	
	\caption{\label{fig:fig7} (a) Distribution of the same-coauthor ratio on KDD Cup dataset; (b) The effects of different attributes.}
\end{figure}

	\textit{KDD Cup}~\cite{roy2013microsoft}: Is the dataset used in the KDD Cup 2013 challenge 1 to address name disambiguation problem. We collect the training data containing 3,739 authors and 123,447 papers, as only the training labels are published. We only use title, organizations, keywords and abstract as features, but ignore coauthor names. As shown in Figure~\ref{subfig:kdd-cup}, the distribution of same-coauthor ratio is extremely skewed. According to Eq.\eqref{eq:same-coauthor-ratio}, same-coauthor ratio equalling 1 means the second similar candidate and the least similar candidate have the same number of same-coauthors with the target pair. In another word, the most similar candidate is significantly different from all the other candidates when only considering the coauthor name features. 
	Thus, 98\% target pairs holding 1.0 same-coauthor ratio means only using the coauthor names can correctly assign 98\% target pairs.
In fact, when considering the coauthor name feature, any baselines including our model can easily achieve approximate 99\% HR@1. Thus, for increasing the difficulty, we ignore coauthor names on this dataset.

\subsubsection{Comparison Methods}


\vpara{Matching Component.}To evaluate the matching performance, we compare feature engineering-based GBDT and three embedding-based models: 

\textit{GBDT}: Is a widely used model to solve KDD Cup 2013 challenge-1~\cite{efimov2013kdd,li2013feature,zhao2013scorecard}. We train a GBDT model to estimate a matching probability between each candidate and the target pair. The extracted features for GBDT are shown in Table~\ref{tb:features}. As the model can directly predict a label for each candidate, it also be used for deciding to assign the most matched candidate to the target pair if its label is 1. 

\textit{Camel}~\cite{zhang2018camel}: Is a representation-based model. Given a triplet $(\langle p,a \rangle,c^+,c^-)$, it first represents $\langle p,a \rangle$ by $p$'s title, and represents $c^+$ and $c^-$ by their identities. Then it calculates the matching scores for both $(\langle p,a \rangle,c^+)$ and $(\langle p,a \rangle,c^-)$, and finally optimizes the difference between their matching scores. 

\textit{HetNetE}~\cite{chen2017task}: Is similar as Camel except that $\langle p,a \rangle$ is represented by all its attributes. 

\textit{GML}~\cite{zhang2018name}: Is a representation-based model to identify whether two papers are written by the same person through optimizing a triplet loss. The model accepts the pre-trained embeddings of all the tokens in a paper as input and output an embedding  for the paper. We represent a person by averaging all his/her papers' embeddings. 

\vpara{Decision Component.}
To evaluate the performance of the decision component, we compare two strategies:

\textit{Threshold}~\cite{gottipati2011linking}: Picks the top matched person whose score is lower than a threshold as NIL, where the threshold is determined as the value when the best accuracy is obtained on a validation set. We use the same matching model as our proposed method to obtain the top matched persons.


\textit{Heuristic Loss}~\cite{clark2016deep}: Unifies the NIL decision and the matching process by incorporating the costs of assigning a paper to a wrong NIL person or assigning an unlinkable paper to a wrong existing person into the loss function of ranking the wrong person before the right person. NIL is inserted as an additional candidate person for each paper. The representations of $p$ and $c$ which are made in the same way as GML, are concatenated as the input of a neural network to produce their matching score. When $c = \text{NIL}$, the representation of $c$ is not included. 

\vpara{Variants of Our Model.} 
We also compare different variants where \PR, \MPR, \sMMR and \sMMPR correspond to the variants in Table~\ref{tb:modelcomponents}. \sMMPC modifies \sMMPR by replacing the triplet loss with the cross-entropy loss, which can be directly used for deciding the assignments. \sRC  trains \sMMPR plus a decision component once. \sJRC jointly trains the two components in \RC.









{\scriptsize \begin{table}[t]
		{\caption{Features extracted for GBDT model. \small{$p$: target paper, $a$: target author in $p$, $c$: candidate person.} \label{tb:features}}} 
		\vspace{-0.08in}
		{\renewcommand{\arraystretch}{1}%
			{
				\setlength{\extrarowheight}{1pt}
				\begin{tabular}{
						@{}c@{ } l@{}}
					\noalign{ \hrule height 1pt}
					\textbf{No.}   & \textbf{Feature description} \\ \hline
					\textbf{1}      &  The number of the papers of $c$\\ 
					 \hdashline
					\textbf{2}      &  The number of the coauthors of $a$ in $p$\\  
					\textbf{3}      &  The number of the coauthors of $c$\\  
					\textbf{4}      &  The number of the same coauthors between $a$ and $c$\\  
					\textbf{5}      &  Ratio of the same coauthors between $a$ and $c$ in $p$'s coauthor names\\ 
					\textbf{6}      &  Ratio of the same coauthors  between $a$ and $c$  in $c$'s coauthor names\\ \hdashline		
					\textbf{7}      &  Frequency of $a$'s affiliation in $c$'s affiliations\\ 
					\textbf{8}      &  Ratio of $a$'s affiliation in $c$'s affiliations\\ 
					\textbf{9}      &  Cosine similarity between $a$'s affiliation and $c$'s affiliations\\ 
					\textbf{10}    &  Jaccards similarity between $a$'s affiliation and $c$'s affiliations\\  \hdashline
					\textbf{11}    &  Distinct number of venues of $c$\\
					\textbf{12}    &  Frequency of $p$'s venue in $c$\\ 
					\textbf{13}    &  Ratio of $p$'s venue in $c$\\ 
					\textbf{14}    &  Cosine similarity between $p$'s venue and $c$'s venues \\ 
					\textbf{15}    &  Jaccards similarity between $p$'s venue and $c$'s venues \\  \hdashline
					\textbf{16}    &  Cosine similarity between $p$'s title and $c$'s titles\\ 
					\textbf{17}    &  Jaccards similarity between $p$'s title and $c$'s titles\\  \hdashline
					\textbf{18}    &  Distinct number of keywords in $c$\\ 	
					\textbf{19}    &  Frequency of $p$'s keywords of $c$\\ 			
					\textbf{20}    &  Ratio of $p$'s keywords in $c$\\ 	 
					\textbf{21}    &  Cosine similarity between $p$'s keywords and $c$'s keywords\\ 
					\textbf{22}    &  Jaccards similarity between $p$'s keywords and $c$'s keywords\\   
					
					\noalign{\hrule height 1pt}
			\end{tabular}}
			
		}
\end{table} }

\subsubsection{Evaluation Settings} 

For each dataset, we randomly sample 20\% persons for testing and divide the rest into training, which results in 45,711 authors for training and 11,427 authors for testing on OAG-WhoIsWho dataset, and 2,991 authors for training and 748 authors for testing on KDD Cup dataset. For each author in both training and testing data, we first sort their papers by the published year in ascending order. Then we choose the latest 20\% papers as the author's unassigned paper and leave 80\% papers as the author profile. 



We first evaluate the matching of the candidate persons to the target pair, and further evaluate the decision of the top matched person as the right person or NIL. 


\vpara{Matching Evaluation.}
For evaluating the matching performance, we sample 10,000 target pairs from the training data. Each target pair paired with its right person composes a positive instance. 
We also sample 9 wrong persons paired with each target paper to compose 9 negative instances. The process results in 90,000 triplets for training.  
For testing, we sample 2,000 target pairs from the test data, where each one is associated with the right person and 19 wrong persons. 

The wrong persons are sampled from the candidates. We follow the name variant strategy in section 3.1 to generate candidates on OAG-WhoIsWho. While for KDD Cup, names are so different that no candidates can be found by simply varying names. Instead, we calculate the Jaro-Winkler similarity between a candidate's name and the target author, and select the candidates whose scores are larger than 0.5 as the wrong persons.

We use Hit Ratio at top k (HR@k) and mean reciprocal rank (MRR) as the metrics for evaluating whether the right person will be ranked at the top among all the candidates. Since there is only one right person for each target pair, HR@k measures the percentage of the candidate lists with the right person ranked before top k. MRR measures the average of reciprocal ranks of the right persons. 
Higher HR@k and MRR indicate better performance.

\vpara{Decision Evaluation.}
We construct the training data for the decision component upon the output of the matching component. Specifically, we also use the 10,000 positive instances  for the matching component  as those for the decision component. Then we extract the target pairs and the corresponding top matched wrong persons to compose the negative instances. 
For testing, in addition to the 2,000 target pairs and the corresponding candidates including the right persons (i.e., positive sample $(\langle p,a \rangle,C)$ with $c^*=c^+$), we extract extra 2,000 target pairs and the corresponding candidates excluding the right persons (i.e., negative sample $(\langle p,a \rangle,C)$ with $c^*=\text{NIL}$). Conveniently, we remove the right person $c^+$ from each positive sample and create a negative sample by the remaining wrong persons.
We count the number of true positive (tp), false negative (fn), true negative (tn) and false positive (fp) samples and then calculate precision, recall and f1:

\beqn{
		\text{tp} &=& |\{ c^* = c^+ \text{  and  } \hat{c} = c^+  \text{  and   } \hat{y} = 1 \}|	 ,\\\nonumber
		\text{fn} &=& |\{ c^* = c^+ \text{  and  }   \hat{y} = 0  \}|	, \\\nonumber
		\text{tn} &=& | \{ c^* = \text{NIL} \text{  and  } \hat{y} =0\} |,	 \\\nonumber
		\text{fp} &=& | \{ c^* = \text{NIL} \text{   and  } \hat{y} =1 \} \cup \\\nonumber
		  &&\{c^*= c^+ \text{  and  } \hat{c} \ne c^+ \text{  and  } \hat{y} = 1\} |，	 \\\nonumber
	}

\noindent where tp is the number of the positive samples, with the right persons ranked at the first (i.e., $\hat{c} = c^+$) and also predicted as the right persons (i.e., $\hat{y}=1$). On the contrary, fn counts the positive samples with $\hat{y}=0$. Notation tn denotes the number of negative samples with the first ranked persons predicted as the wrong persons (i.e., $\hat{y}=0$), while fp counts the negative samples with $\hat{y}=1$ and also counts the positive samples with the wrong persons ranked at the first (i.e., $\hat{c} \ne c^+$) but still predicted as the right persons (i.e., $\hat{y}=1$).
Since we aim at assigning the target pair to an existing right person and also assigning it to NIL if there is no right person, we  calculate precision and recall for both the cases with $c^* = c^+$ and $c^* = \text{NIL}$:

\beqn{
	c^* = c^+\!\!\!&:& \!\!\!\text{Pre.} = \frac{\text{tp}}{\text{tp + fp}},   \quad \text{Rec.} = \frac{\text{tp}}{\text{tp + fn}}; \\ \nonumber
	c^* = \text{NIL} \!\!\!&:&\!\!\!\text{Pre.} = \frac{\text{tn}}{\text{tn + fn}},   \quad \text{Rec.}  = \frac{\text{tn}}{\text{tn + fp}}.
}

\subsubsection{Implementation Details}

We divide the attributes of a paper into two fields: coauthor names and other attributes including title, abstract, organizations and keywords, as coauthor names have no literal or semantic overlaps with other attributes.
We pre-train an embedding for each author name and each word. Specifically, we use Word2Vec to train an embedding for an author name in the context of all the coauthors' names in a paper, and train an embedding for a word in the context of all the other occurred words in title, keywords, venue and affiliation.
We set the dimension of the embedding as 100. To enable matrix operation, for each paper or candidate person, we restrict the maximal number of author names to 100, the maximal number of words to 500, and the maximal number of papers published by each person to 100. 

The hyper-parameters of the RBF kernel functions are set the same as~\cite{xiong2017end}. We use 11 RBF kernels, with the hyper-parameters $\mu$=$\{1, 0.9, 0.7, 0.5, 0.3, 0.1, -0.1, -0.3, -0.5, -0.7, -0.9\}$  and $\sigma$=$\{10^{-3}, 0.1, 0.1, 0.1, 0.1, 0.1, 0.1, 0.1, 0.1, 0.1, 0.1\}$.

Function $g$ in Eq.\eqref{eq:rankloss} is instantiated as a 3-layer MLP followed by a ReLU function which transforms a similarity embedding $\phi(\langle p,a \rangle, c)$ into a 1-dimensional score. Function $h$ in Eq.\eqref{eq:decision} is also a 3-layer MLP which transforms a $\phi(\langle p,a \rangle, c)$ into 2-dimensional classification probabilities.



\subsection{Performance Analysis}

\begin{table}
	\newcolumntype{?}{!{\vrule width 1pt}}
	\newcolumntype{C}{>{\centering\arraybackslash}p{3.5em}}
	\caption{
		\label{tb:ranking_perforamance} Performance of the matching results (\%).
		\normalsize
	}
	\centering  \footnotesize
	\renewcommand\arraystretch{1.0}
	\begin{tabular}{@{}c@{~}?*{1}{m{0.6cm}m{0.6cm}m{0.6cm}?}*{1}{m{0.6cm}m{0.6cm}m{0.6cm}}}
		\toprule
		\multirow{2}{*}{Model}
		&\multicolumn{3}{c?}{OAG-WhoIsWho}
		&\multicolumn{3}{c@{}}{KDD Cup} 
		
		\\
		\cmidrule{2-4} \cmidrule{5-7} 
		& {HR@1} & {HR@3} & {MRR} & {HR@1} & {HR@3} & {MRR}\\
		\midrule
		Camel
		&41.20&62.00& 55.00
		& 44.62 & 67.19  & 59.44	
		\\ 
		HetNetE
		&46.00&67.00&60.24	 
		& 51.06 & 77.44  &   66.41	 
		\\
		GML
		&70.87&94.53&82.59
		&72.13&95.34&82.90	
		\\
		GBDT
		&87.30&98.10&92.71
		&84.18&92.09&89.59
		\\ \midrule
		\PR
		& 86.20 & 96.40 & 92.20
		& 91.12& 95.72 & 93.73
		\\
		\MPR
		& 88.00 & 98.75 & 93.25	
		&-&-&-
		\\
		\MMR
		& 89.45 & 98.40 & 93.82
		& 91.45 & 95.80 & 94.03
		\\
		\midrule
		\textbf{\RC}
		&  90.45 &   98.30 &  94.46
		& 92.10 & 96.35& 94.66
		\\
		\textbf{\JRC}
		& \textbf{91.10} &  \textbf{98.45} &  \textbf{94.86}
		& \textbf{92.60} &  \textbf{96.71} &  \textbf{94.95}
		\\

		\bottomrule
		
	\end{tabular}
	
\end{table}

\subsubsection{Matching Performance}

\vpara{Overall Matching Performance.}
Table~\ref{tb:ranking_perforamance} shows the matching performance of the proposed model, the model variants and the comparison methods on the two datasets OAG-WhoIsWho and KDD Cup. In terms of HR@1, the proposed \sJRC achieves 3.80\% to 49.90\% improvement over all the baseline methods. 

Camel, HetNetE and GML are all  representation-based deep learning models, which can capture the soft/semantic matches, but they will dilute the effect of the exact matches of tokens due to the global representations of the papers and persons.   
Among the three models, HetNetE uses all the attributes of a paper rather than the single title to represent a paper, which achieves better performance than Camel. Camel and HetNetE represent the candidate persons only based on their identities. Thus they suffer from the sparsity issue, i.e, the embeddings of the persons cannot be trained accurately if they publish few papers. GML avoids the sparsity issue through representing persons by their published papers. However, it is difficult to directly compare the embeddings of a long text (i.e, all the papers of a candidate person) and a short text (i.e., a target paper). 

In the name disambiguation problem, the exact matches between tokens especially the matches between coauthor names are more important than the soft matches, thus although GBDT only captures the exact matches, it performs better than the representation-based models. The proposed interaction-based matching component in \sRC captures both the exact and the soft matches through comparing local representations of each token pairs instead of comparing the global representations of papers and persons. Specifically, the kernel aggregation function used in  the matching component summarizes a frequency distribution of the exact matches and different kinds of soft matches, which can't dilute the effect of extract matches by the other soft matches. 
Thus, the proposed matching component performs better than  all the comparison methods. 

Compared with \RC, the performance of \sJRC is further improved, as the decision component gives additional feedbacks to supervise  the ranking of the matching component.
The result indicates that through jointly fine-tuning of the two components, the errors of the matching component can be reduced.

\begin{figure}[t]
	\centering
	\includegraphics[width=0.5\textwidth]{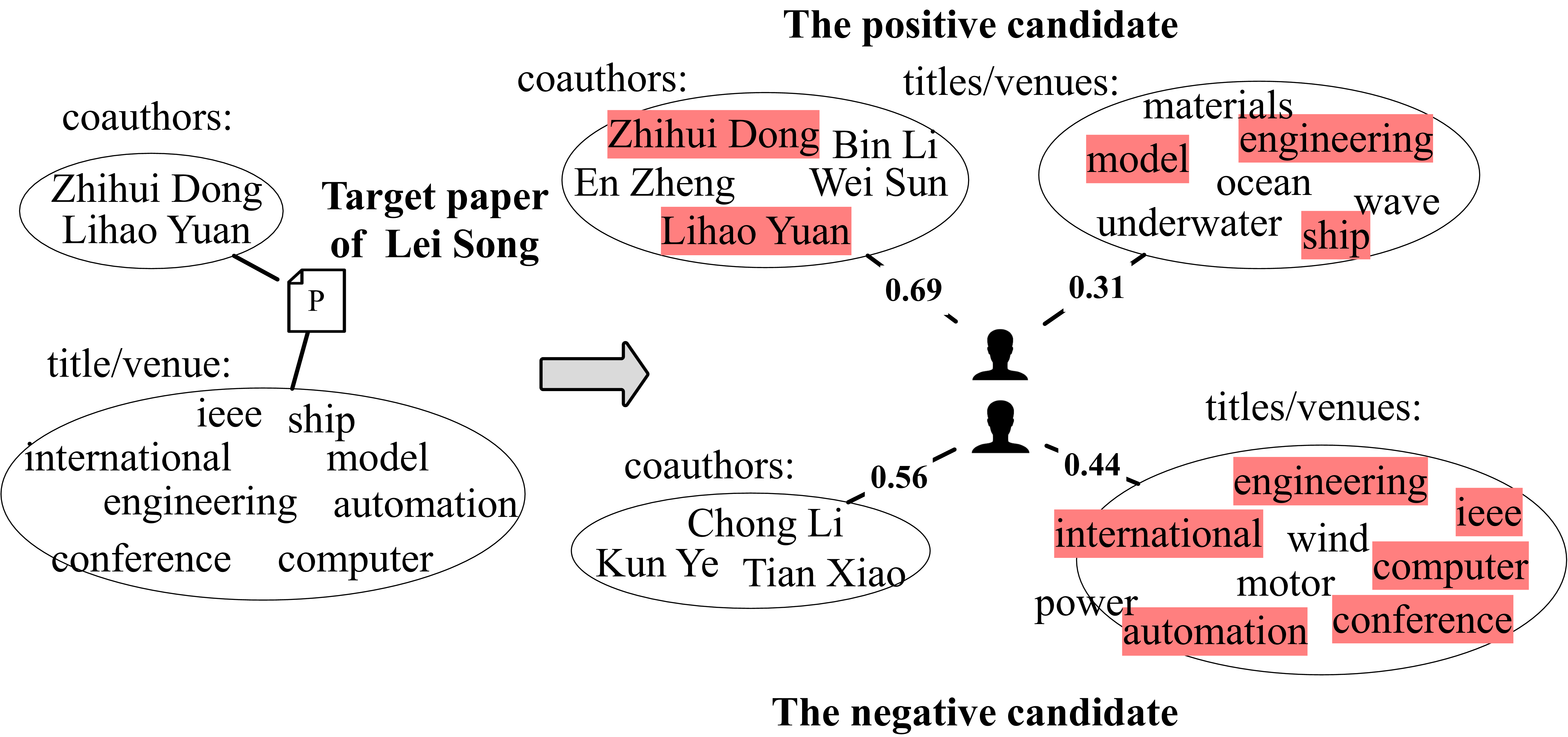}
	\caption{\label{fig:multi-field-case} Case study of multi-field effect.}
\end{figure}
\begin{figure}[t]
	\centering
	\includegraphics[width=0.5\textwidth]{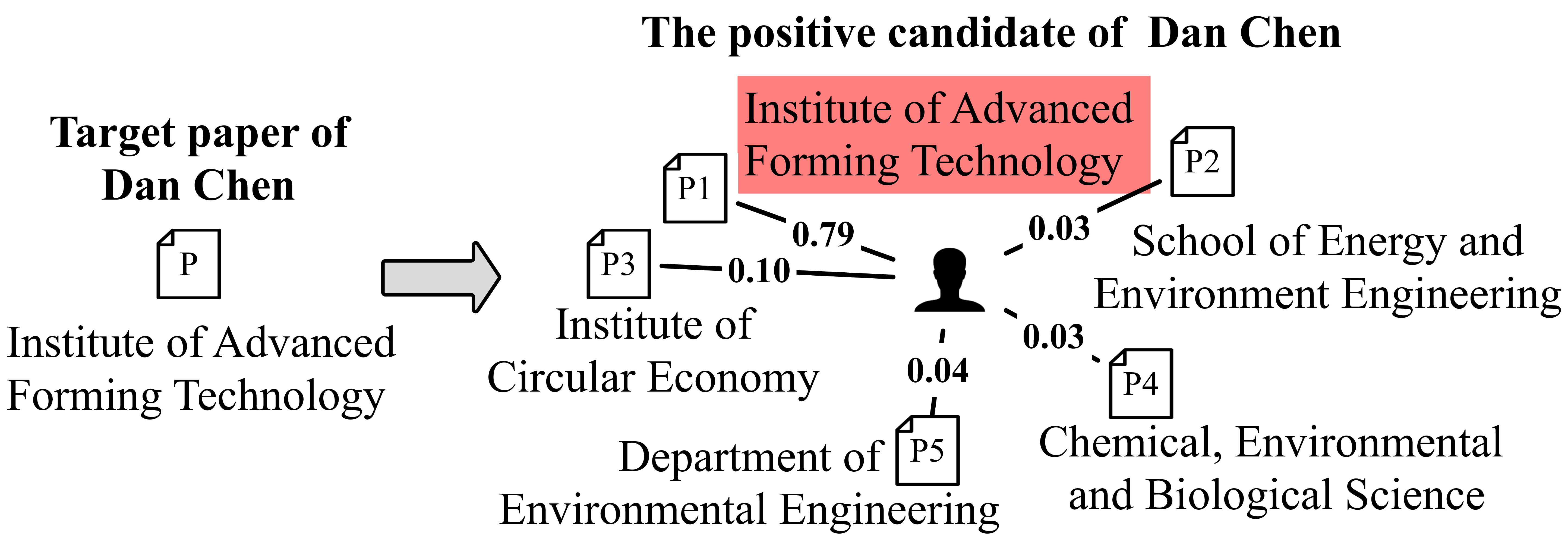}
	\caption{\label{fig:multi-instance-case} Case study of multi-instance effect.}
\end{figure}

Comparing the results on the two datasets, we can see that the advantage of our model over the feature engineering-based GBDT is much more significant on KDD Cup (+8.42\% in HR@1) than OAG-WhoIsWho (+3.80\% in HR@1). Since coauthor features are not used on the KDD Cup, the results indicate that \sRC can better capture the semantics of the attributes except coauthor names.

\vpara{Multi-field Effect.}
We conduct an ablation study to analyze the effects of different modeling strategies on the matching component. Since only one field  is used on the KDD Cup dataset, we analyze the effect of multi-fields on the OAG-WhoisWho dataset. From Table~\ref{tb:ranking_perforamance}, we can see that \sMPR performs better than \sPR (improving 1.8\% in terms of HR@1), which indicates that it is necessary to build the interaction-based models for different attributes separately and distinguish their effects. 

We also investigate the effects of different fields by removing coauthor names and other attributes respectively based on the model \RC. The experimental results in Figure~\ref{subfig:name_attr} show that removing either coauthor names or other attributes performs significantly worse (-5.80\%-7.05\%, HR@1) than \RC, which indicates that both coauthor names and other attributes impact the performance obviously. What's more, removing names is comparable to removing other attributes, which indicates that coauthor names are more important than all the other attributes on the task of name disambiguation.

\vpara{Multi-instance Effect.} 
Table~\ref{tb:ranking_perforamance} also shows that on OAG-WhoisWho, \sMMR performs better than \sMPR (+1.45\% in terms of HR@1), which demonstrates the strength of distinguishing different papers of a person. HR@1 of \sMMR is further improved by 1.00\% if we combine the profile model \sMPR and the multi-instance model \sMMR as \RC. The result indicates that both the global similarity between the target paper and a candidate's whole profile, and the local similarities between the target paper and each paper of a candidate take effects on matching performance. The results on KDD Cup also present the advantages of multi-instances.

\begin{table*}
	\newcolumntype{?}{!{\vrule width 1pt}}
	\newcolumntype{C}{>{\centering\arraybackslash}p{3.5em}}
	\caption{
		\label{tb:classification_perforamance} Performance of the decision results (\%).
		\normalsize
	}
	\centering  \footnotesize
	\renewcommand\arraystretch{1.0}
	\begin{tabular}{@{}c@{~}?*{1}{m{0.6cm}m{0.6cm}m{0.6cm}?}*{1}{m{0.6cm}m{0.6cm}m{0.6cm}?}*{1}{m{0.6cm}m{0.6cm}m{0.6cm}?}*{1}{m{0.6cm}m{0.6cm}m{0.6cm}}}
		\toprule
		\multirow{3}{*}{Model}
		&\multicolumn{6}{c?}{OAG-WhoIsWho}
		&\multicolumn{6}{c@{}}{KDD Cup} 
		
		\\
		\cmidrule{2-4} \cmidrule{5-7}  
		\cmidrule{8-10}  \cmidrule{11-13} 
		
		&\multicolumn{3}{c?}{Samples with $c^*=c^+$}
		&\multicolumn{3}{c?}{Samples with $c^*=\text{NIL}$}
		&\multicolumn{3}{c?}{Samples with $c^*=c^+$}
		&\multicolumn{3}{c@{}}{Samples with $c^*=\text{NIL}$}		
		\\
		\cmidrule{2-4} \cmidrule{5-7}  \cmidrule{8-10}  \cmidrule{11-13} 
		& {Pre.} & {Rec.} & {F1} & {Pre.} & {Rec.} & {F1}
		& {Pre.} & {Rec.} & {F1} & {Pre.} & {Rec.} & {F1}
		\\
		\midrule
		GBDT
		& 82.87 & 72.40 & 77.28  	
		& 75.39 & 85.04 & 79.98  
		& 83.64 & 71.64 & 77.17 
		& 75.20 & 85.98 & 80.23
		\\ 
		
		Threshold
		& 79.33 & 57.60 & 66.38  	
		& 66.47 & \textbf{84.07} & 74.24  
		& 74.89 & 71.00 & 72.90 
		& 72.43 & 76.20 & 74.27
		\\
		Heuristic Loss
		& 71.79 & 78.40 & 74.95  	
		& 76.21 & 69.20 & 72.54  
		& 85.14 & 69.60 & 76.59 
		& 74.29 & 87.85 & 80.50
		\\
		\MMPC 
		& 79.42 & 82.33 & 80.85  	
		& 81.66 & 78.67 & 80.14 
		& 89.60 & 82.79 & 86.06 
		& 86.15 & 88.05 & 87.09
		\\ 	
		\midrule
		\textbf{\RC}
		&79.53&89.87&84.38
		&88.35&76.87&82.21
		&88.44& \textbf{86.20}&87.31 
		&\textbf{86.54}&88.73&87.62
		\\
		\textbf{\JRC}
		&\textbf{82.47} & \textbf{90.33} & \textbf{86.22}
		&\textbf{89.31} & 80.80 & \textbf{84.84}
		&\textbf{89.87}&85.73&\textbf{87.75}
		&86.36&\textbf{90.33}&\textbf{88.30}
		\\
		
		\bottomrule
		
	\end{tabular}
	
\end{table*}


\vpara{Interpretability of the Matching Component.}
We present some cases in Figure~\ref{fig:multi-field-case} and Figure~\ref{fig:multi-instance-case} to demonstrate the interpretability of the proposed matching component. From Figure~\ref{fig:multi-field-case}, we can see that although the number of the matched tokens between the target paper and the positive candidate person is less than that of the negative candidate person, the matched coauthors are more important than the matched words in titles and venues, because the attention $\alpha$ learned by our model for the matched coauthors on the positive candidate is 0.69, comparing with 0.31 learned for the matched titles and venues. And the attention learned on the negative candidate also emphasizes the matched coauthors.  \sRC distinguishes different fields' effects by the attention, thus it can correctly identify the positive candidate, while the basic profile model \sPR wrongly returns the negative candidate as the most matched candidate, as it treats the matches in all the fields equally.

In Figure~\ref{fig:multi-instance-case}, we present the affiliation of ``Dan Chen" in both the target paper and the positive candidate. It is shown that a paper of the positive candidate has the same affiliation with the target paper, and the corresponding attention $\beta$ learned by our model for the paper is 0.79, while the values of $\beta$ learned for other papers are much smaller than this paper. \sRC distinguishes different papers' effects, thus it can correctly identify the positive candidate, while the basic profile model \sPR treats the matches in all the papers equally, which dilutes the effects of similar papers by the other irrelevant papers. The learned attentions for different fields and different papers both demonstrates the interpretability of the proposed matching component.

\vpara{Matching Performance on Different Scenarios.}
We conduct additional experiments on the matching performance of different baselines and \sRC with different same-coauthor ratios on OAG-WhoIsWho dataset and present the results in Figure~\ref{fig:proportion_coauthor}. ﻿We can see that HR@1 of the embedding-based models, i.e. Camel, HetNetE, GML and \sRC  drop more slightly (drops from 6.63\% to 18.73\%) than feature-engineering based GBDT (drops more than 29.69\%) when the same-coauthor ratio decreases from 1.0 to 0.1. This indicates that the embedding-based model can better capture the semantic matches when the coauthor features are week. 
Especially when the same-coauthor ratio is less than 0.1, the performance
gap between \sRC and GBDT is significantly more than 16\%. The result indicates
that \sRC is more suitable to tackle the hard cases, i.e.
the cases that are hardly predicted by similar coauthors.


\begin{table}
	\newcolumntype{?}{!{\vrule width 1pt}}
	\newcolumntype{C}{>{\centering\arraybackslash}p{3.5em}}
	\caption{
		\label{tb:effciency} Average time cost(ms) of assigning each target pair.
		\normalsize
	}
	\centering 
	\renewcommand\arraystretch{1.0}
	\begin{tabular}{cccc}
		\toprule
		Model & Feature Preparing & Matching & Decision\\
		\midrule
		GBDT
		& 183.34 & - & 3.61
		\\ 
		\midrule
		\RC
		& 260.45 & 76.12 & 6.34
		\\
		\bottomrule
		
	\end{tabular}
	
\end{table}

\subsubsection{Decision Performance}

Table~\ref{tb:classification_perforamance} shows the final decision performance of the proposed model and the comparison methods. Comparing with other methods, in terms of F1, the proposed joint model \sJRC achieves 1.69\%-19.84\% improvement on the samples with $c^*=c^+$ and 1.21\%-14.03\% improvement on the samples with $c^*=\text{NIL}$. We evaluate the results on both of the samples as we aim at not only assigning the target papers to the right persons if they exist, but also assigning them to NIL if the right persons do not exist.
The problem in this paper is not merely a matching or a classification decision problem, but can be solved by firstly matching each candidate to the target paper $p$ and then deciding whether the top matched person is right or not. Thus, we need to not only keep the relevant order within each candidate list, but also globally distinguish all the positive pairs from all the negative pairs. 

GBDT and \sMMPC only aim to optimize the global positions of all the $(\langle p,a \rangle,c)$ pairs, but ignore the relative order within each candidate list. 
Although the globally predicted probabilities can be used to compare the candidates of each target paper, the relative order is not directly optimized, leading to a lot of mistakes in the final results.
Threshold can be viewed as a global optimization model, but merely uses a heuristic threshold to distinguish different complicated cases. 
Heuristic Loss incorporates the costs related to NIL into the original loss of ranking the wrong persons before the right persons, but it suffers from the heuristically configured weights of different costs.

\sRC first estimates the matching probability of each candidate to the target pair and then decides the top matched candidate. This two-step strategy which is widely adopted in entity linking~\cite{mcnamee2010hltcoe,ratinov2011local} is proved to be effective. 
Compared with \RC, the performance of \sJRC is further improved, as some of the wrongly-predicted instances are gradually represented better to generate accurate similarity embeddings by the iteratively refined matching component, which will finally increase the number of rightly predicted instances.
The result demonstrates that the errors of the decision component can be reduced through jointly fine-tuning of the two components.

\begin{figure}
	\centering
	\subfigure[Matching component.]{\label{subfig:matching_convergence}
		\includegraphics[width=0.22\textwidth]{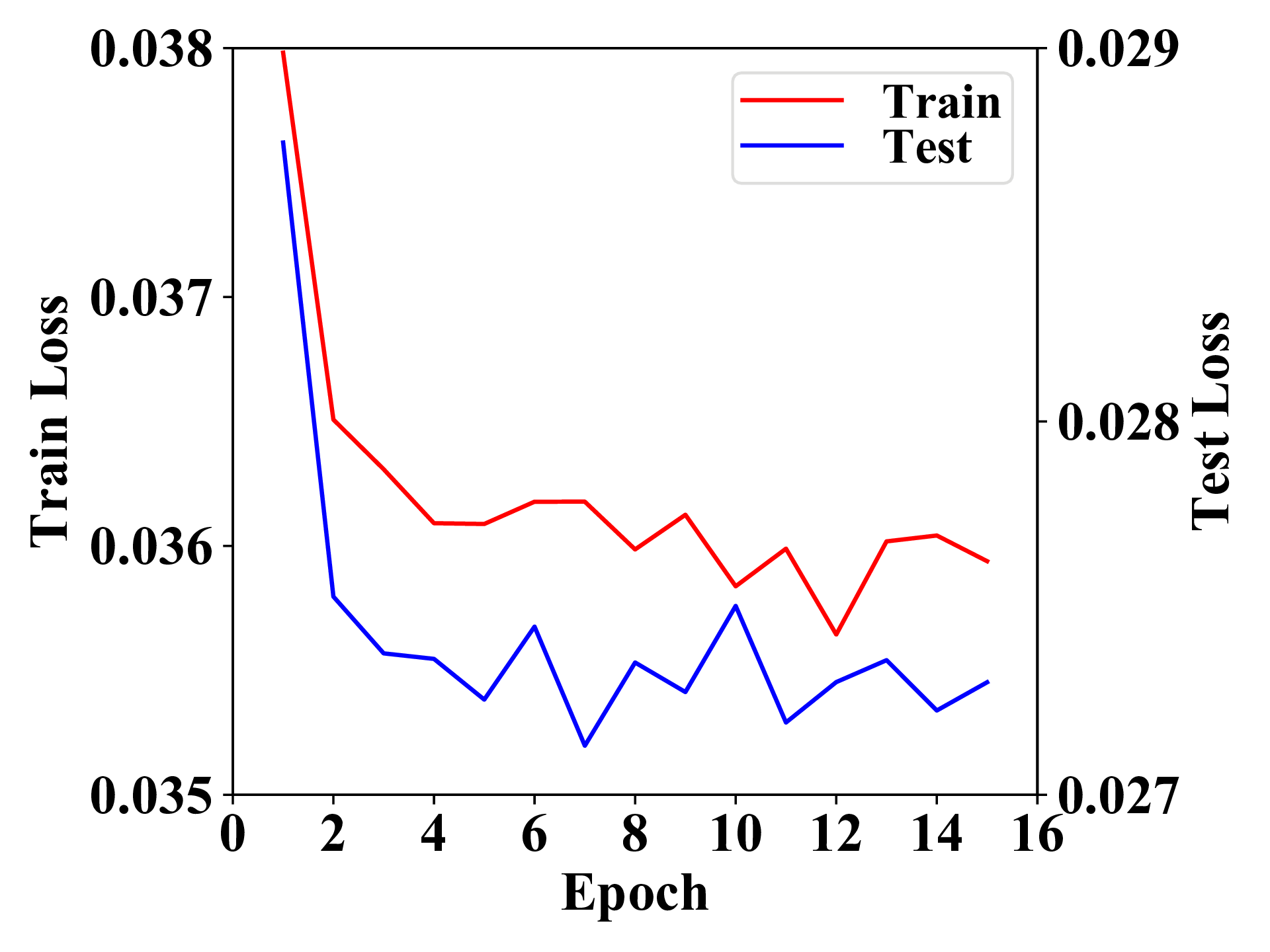}
	}
	\subfigure[Decision component.]{\label{subfig:decision_convergence}
		\includegraphics[width=0.22\textwidth]{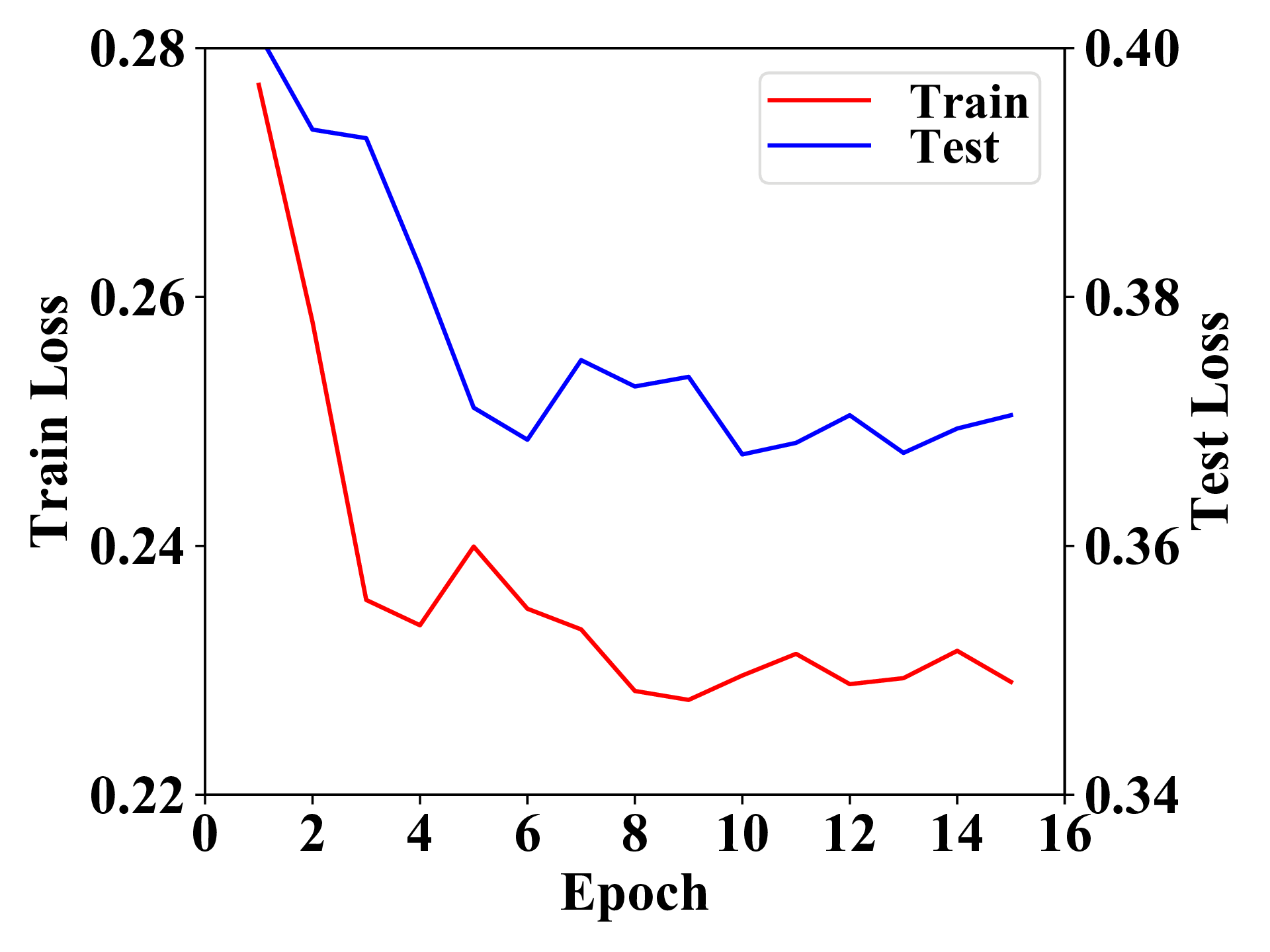}
	}
	
	\caption{\label{fig:convergence} Convergence Analysis.}
\end{figure}

\vpara{Convergence Analysis.}
We plot the train/test loss of the matching component and the decision component with the increase of the joint training epochs. The results in Figure~\ref{fig:convergence} show that the performance of the two components both decrease sharply at the beginning of the joint training and then gradually change stable, which indicate the convergence of \JRC.

\subsection{Online Deployment on AMiner}
\label{sec:online}
Table~\ref{tb:effciency} presents the average time cost of assigning each target paper by the proposed \sRC model and the best baseline GBDT. 
We implement the experiments by Tensorflow and run the code on an Enterprise Linux Server with 40 Intel(R) Xeon(R) CPU cores (E5-2640 v4 @ 2.40GHz  and 252G memory) and 1 NVIDIA Tesla V100 GPU core (32G memory).
Since GBDT is a classification model without the matching component, we only present the cost of the decision process, which uses the label of the top predicted candidate as the predictive result. From Table~\ref{tb:effciency}, we can see that \sRC is about 1.83$\times$  slower than GBDT, which is mainly determined by the feature preparing process. Although \sRC performs much better than GBDT on both the ranking and the decision performance, from Figure~\ref{fig:proportion_coauthor}, we can see for about 62.17\% easy samples, i.e., the target pairs with the same-coauthor ratio larger than 0.9, the ranking performance of GBDT is comparable to \RC, where the ranking performance directly determines the final decision performance of the top-1 candidates. Thus, to improve the online assignment efficiency meanwhile keeping the assignment performance, for each target pair, if its same-coauthor ratio is larger than 0.9, we directly apply GBDT to perform paper assignment, otherwise we apply \sRC to complete the task. 

In addition, the online candidate selection is a little different from the offline name variant strategy explained in~\secref{sec:overview}. To improve the recall of the online predicting as much as possible, we adopt ElasticSearch\footnote{https://www.elastic.co} to perform fuzzy search for similar candidates with each target author. Compared with this online fuzzy strategy, the offline candidate selection is more strict, as for annotating high-quality name disambiguation dataset, the simple name variant strategy can already produce enough challenging candidates. However, the fuzzy strategy may result in too many noisy candidates, which increase annotation efforts.


\begin{figure}
	\centering
	
	\subfigure[A $c^*=c^+$ case.]{\label{subfig:normal_case}
		\includegraphics[width=0.48\textwidth]{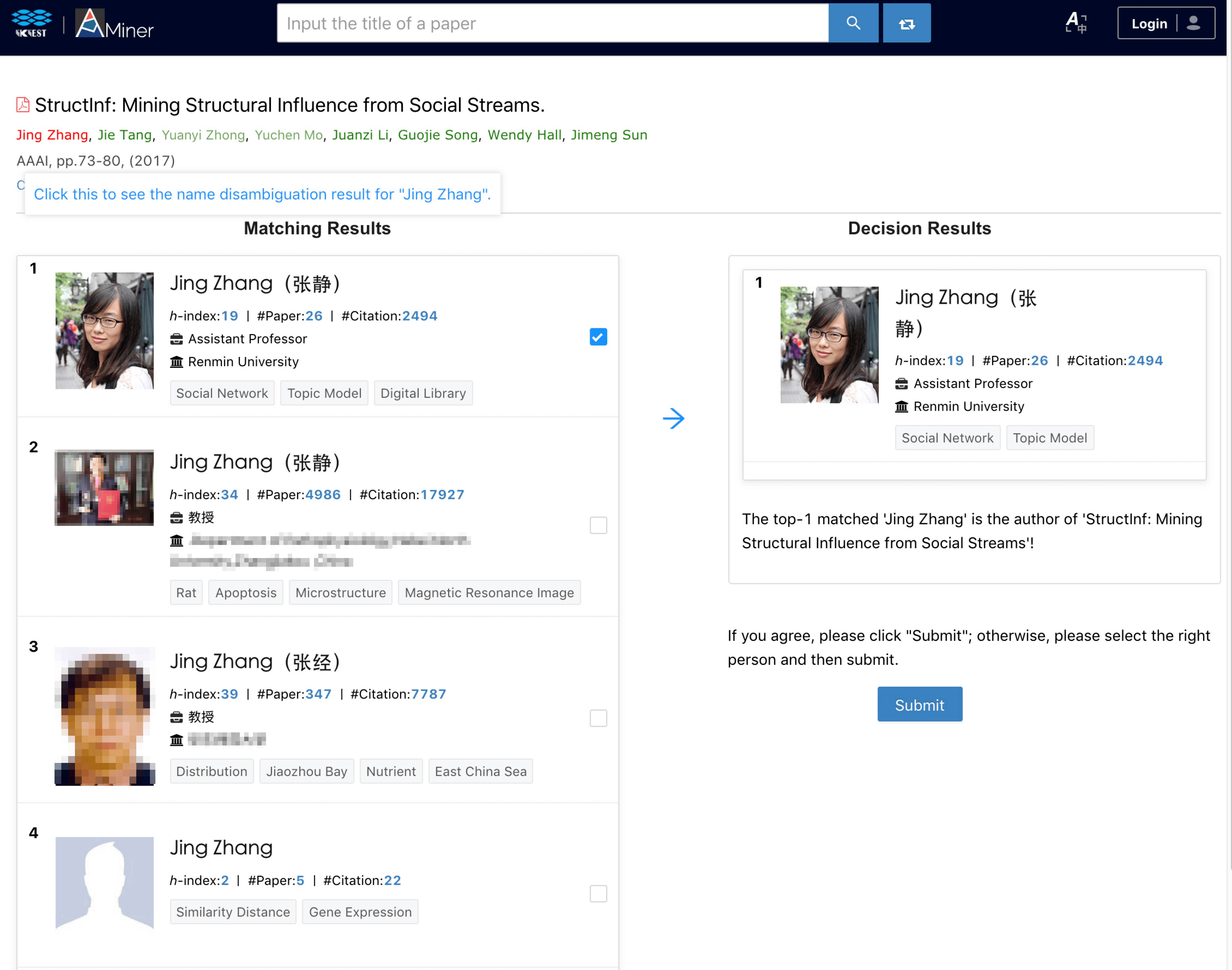}
	}
	\subfigure[A $c^*=\text{NIL}$ case.]{\label{subfig:nil_case}
		\includegraphics[width=0.48\textwidth]{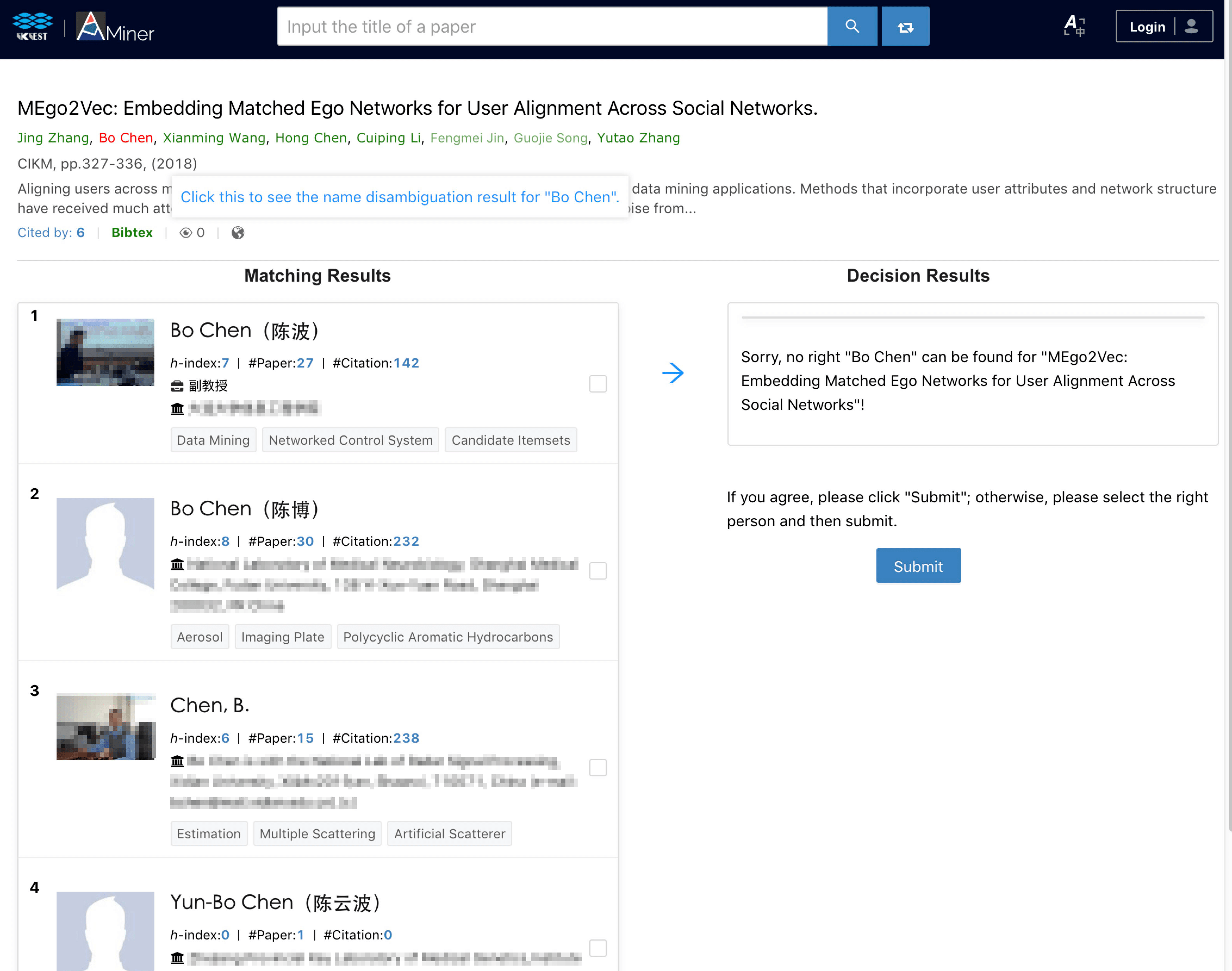}
	}
	
	\caption{\label{fig:demo} A demo of disambiguation on the fly in AMiner.}
\end{figure}

We develop a demo of disambiguation on the fly in AMiner\footnote{http://na-demo.aminer.cn/}, and show two screenshots of the demo in Figure~\ref{fig:demo}. In the demo, users are allowed to search a paper by its title, then select the expected paper and click one author name to see the disambiguation results of the paper with the current name.  Under the selected paper, we present the most matched candidates by the trained matching component in \sRC on the left, and show the decision result of the assigned person by the trained decision component in \sRC on the right. Figure~\ref{subfig:normal_case} shows a case with $c^*=c^+$. We can see that our model can correctly match ``Jing Zhang" from Renmin University for the author ``Jing Zhang" in the paper ``StructInf: Mining Structural Influence from Social Streams" at the top and then decide the top matched one as the final assigned person. 
Figure~\ref{subfig:nil_case} shows a case with $c^*=\text{NIL}$. Since ``Bo Chen" of the paper ``MEgo2Vec: Embedding Matched Ego Networks for User Alignment Across Social Networks" is a postgraduate student whose profile has not been established by AMiner, none of the existing ``Bo Chen" should be assigned to the paper. Our model correctly assigns NIL to this case.
Besides, since errors are still inevitable, we allow the users to provide feedback to our decision results. Specifically, users are allowed to directly ``submit" the result if they agree with it, otherwise, they can choose another right person from the top matched persons. The feedback can be simply regarded as new training instances to update the decision performance at each step of the joint training.

\section{Related Work}
\label{sec:related}

This paper is related to the problems of name disambiguation from scratch, author identification and entity linking.

 \vpara{Name Disambiguation from Scratch.} Much effort has been made to disambiguate names from scratch defined as: given a set of papers written by the authors with similar name, it targets at partitioning all the papers into several disjoint clusters, with each of them corresponds to a real person. Existing work firstly represent papers by traditional feature engineering methods~\cite{chen-martin-2007-towards, huang2006efficient,tang2012unified,wang2010constraint,wang2011adana} or embedding  models~\cite{qiao2019unsupervised,wang2020author,zhang2017name,zhang2018name}  and then adopt a clustering algorithm such as hierarchical agglomerative clustering~\cite{chen-martin-2007-towards, qiao2019unsupervised,wang2020author,zhang2017name,zhang2018name}, K-means~\cite{wang2010constraint, chen2019toward}, DBSCAN~\cite{huang2006efficient} or semi-supervised clustering~\cite{louppe2016ethnicity} to partition these papers. Embedding models  further include graph auto-encoder~\cite{zhang2018name}, heterogeneous GCN~\cite{qiao2019unsupervised} and adversarial representation learning~\cite{wang2020author}. 
 Continuous name disambiguation is formalized differently from the above problem, thus it can not be solved by the above methods.

\vpara{Author Identification.} Several works devote to anonymous author identification for a paper, which assume the authors of the target paper are unknown in a double-blind setting. For example, Chen et al.~\cite{chen2017task} and Zhang et al.~\cite{zhang2018camel} both optimize the difference between the right and the wrong authors. However, their models cannot be applied to unseen authors in the training set, as they only consider the identities of the authors. While we model authors' profiles, which do not depend on authors' identities. KDD Cup 2013 held an author identification challenge to solve the similar problem. However, the situation that no right person exists was not considered and all the participations devoted to feature-engineering methods~\cite{efimov2013kdd,zhao2013scorecard}.

\vpara{Entity Linking.} Entity linking aims at linking the mentions extracted from the unstructured text to the right entities in a knowledge graph~\cite{shen2014entity}. 
Feature-based ~\cite{lehmann2010lcc} or neural models such as skip-gram~\cite{yamada2016joint}, autoencoder~\cite{he2013learning}, CNN~\cite{sun2015modeling}, LSTM~\cite{Kolitsas2018ACL} are proposed to calculate the similarity between the context of a mention and a candidate entity. 
The NIL problem is widely studied in entity linking. The main solutions usually include the NIL threshold methods~\cite{gottipati2011linking,shen2013linking} predicting the mention as unlinkable if the score of the top ranked entity is smaller than a NIL threshold, the classification methods~\cite{mcnamee2010hltcoe,ratinov2011local} which predict the unlinkable mentions by a binary classifier, and the unified models incorporating unlinkable mention prediction process into entity matching process~\cite{clark2016deep,han2011generative}. Different from above, we jointly train the NIL decision model and the candidate matching model to boost both of their performance.

\section{Conclusion}
\label{sec:con} 
This paper presents the first attempt to formalize and solve the problem of name disambiguation on the fly by considering different cases of assignments, in particular when a paper cannot be assigned to any existing persons in the system. We propose a novel joint model that consists of a matching component and a decision component, where a multi-field multi-instance interaction-based model is trained to match the candidates to each target paper, and then a classification decision model is trained to decide whether to assign the top matched candidate to the target paper or not. Through reinforcement joint fine-tuning, the two components can bootstrap each other and self-correct some of their errors. The experimental results on the recent largest dataset for name disambiguation demonstrate that the proposed model performs significantly better than state-of-the-art baseline methods. The model has already been deployed on AMiner to disambiguate the online papers.

\section*{Acknowledgments}
This work is supported by National Key R\&D Program of China (No.2018YFB1004401) and NSFC  (No.61532021, 61772537, 61772536, 61702522)

\bibliographystyle{abbrv}
\bibliography{reference}

%
\vspace{-0.5in}
\begin{IEEEbiography}[\InsertBoxL{-6}{\includegraphics[width=0.7\textwidth,height=3in,clip,keepaspectratio]{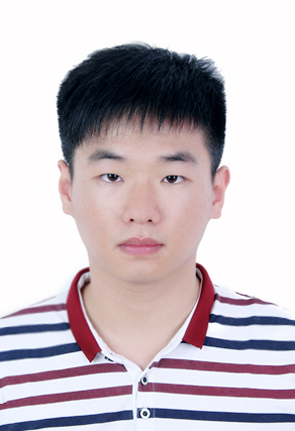}}]{Bo Chen}
	is a graduate student in Information School, Renmin University of China. His research interests include data integration and knowledge graph mining and he recently focuses on entity disambiguation in academic networks.
\end{IEEEbiography}

\vspace{-0.6in}
\begin{IEEEbiography}[\InsertBoxL{-6}{\includegraphics[width=0.7\textwidth,height=3in,clip,keepaspectratio]{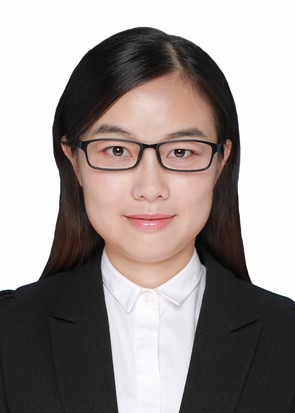}}]{Jing Zhang}
received the master and  PhD degree from the Department of Computer Science and Technology, Tsinghua University. She is an assistant professor in Information School, Renmin University of China. Her research interests include knowledge graph constructing and mining. 
\end{IEEEbiography}

\vspace{-0.6in}
\begin{IEEEbiography}[\InsertBoxL{-5}{\includegraphics[width=0.7\textwidth,height=3in,clip,keepaspectratio]{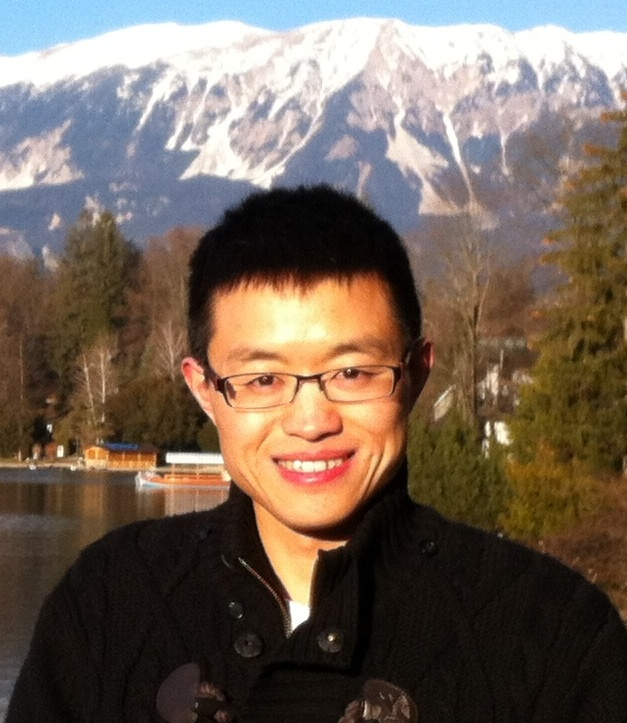}}]{Jie Tang} is a professor in the Department of Computer Science and Technology, Tsinghua University. His research interests include data mining, social network, and machine learning. He was honored with the SIGKDD Test-of-Time Award in Applied Data Science.
\end{IEEEbiography}

\vspace{-0.6in}
\begin{IEEEbiography}[\InsertBoxL{-6}{\includegraphics[width=0.7\textwidth,height=3in,clip,keepaspectratio]{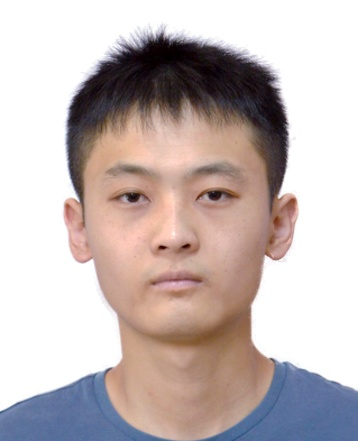}}]{Lingfan Cai}
	is a graduate student in Information School, Renmin University of China. His research interests includes name disambiguation in academic networks.
\end{IEEEbiography}

\vspace{-0.6in}
\begin{IEEEbiography}[\InsertBoxL{-6}{\includegraphics[width=0.7\textwidth,height=3in,clip,keepaspectratio]{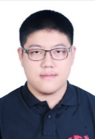}}]{Zhaoyu Wang} is a graduate student in the School of Computer Science and Technology, Anhui University. His research interests include name disambiguation in patent datasets.
\end{IEEEbiography}

\vspace{-0.6in}
\begin{IEEEbiography}[\InsertBoxL{-6}{\includegraphics[width=0.7\textwidth,height=3in,clip,keepaspectratio]{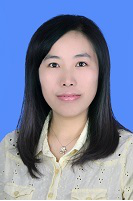}}]{Shu Zhao} received her PhD degree from Anhui University in 2007. She is a professor in the  School of Computer Science and Technology of Anhui University. Her current research interests include social network, machine learning and granular computing.
\end{IEEEbiography}

\vspace{-0.6in}
\begin{IEEEbiography}[\InsertBoxL{-6}{\includegraphics[width=0.7\textwidth,height=3in,clip,keepaspectratio]{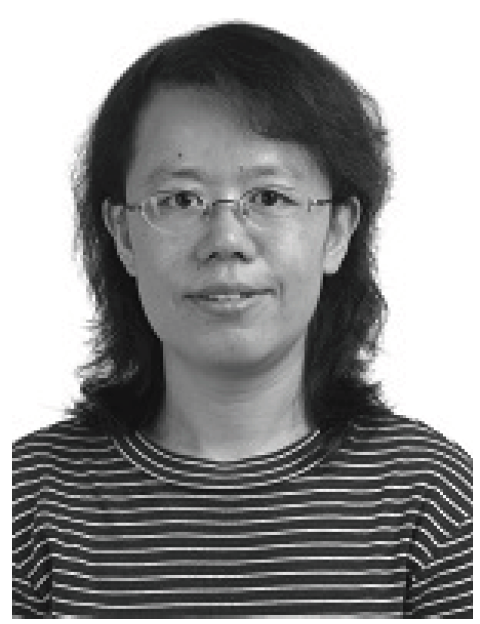}}]{Hong Chen} 
	received PhD degree from the Institute of Computing Technology, CAS. She is a professor of Renmin University of China. Her research interests include data privacy, big data management, and data analysis based on new hardwares. She is a distinguished member of the CCF.
\end{IEEEbiography}

\vspace{-0.6in}
\begin{IEEEbiography}[\InsertBoxL{-6}{\includegraphics[width=0.7\textwidth,height=3in,clip,keepaspectratio]{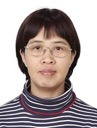}}]{Cuiping Li} 
	 received the PhD degree from the Institute of Computing Technology, CAS. She is currently a professor of Renmin University of China. Her current research interests include database systems, social network analysis, and data mining.
\end{IEEEbiography}




\end{document}